\documentclass[10pt]{article}

\PassOptionsToPackage{unicode}{hyperref}
\PassOptionsToPackage{hyphens}{url}

\usepackage{amsmath,amssymb}
\usepackage{iftex}

\ifPDFTeX
  \usepackage[T1]{fontenc}
  \usepackage[utf8]{inputenc}
  \usepackage{textcomp} % provide euro and other symbols
\else % if luatex or xetex
  \usepackage{unicode-math} % this also loads fontspec
  \defaultfontfeatures{Scale=MatchLowercase}
  \defaultfontfeatures[\rmfamily]{Ligatures=TeX,Scale=1}
\fi

\usepackage{lmodern}
\usepackage{booktabs}
\usepackage{comment}

\ifPDFTeX\else
\fi

\IfFileExists{upquote.sty}{\usepackage{upquote}}{}

\IfFileExists{microtype.sty}{% use microtype if available
  \usepackage[]{microtype}
  \UseMicrotypeSet[protrusion]{basicmath} % disable protrusion for tt fonts
}{}

\makeatletter
\@ifundefined{KOMAClassName}{% if non-KOMA class
  \IfFileExists{parskip.sty}{%
    \usepackage{parskip}
  }{% else
    \setlength{\parindent}{0pt}
    \setlength{\parskip}{6pt plus 2pt minus 1pt}}
}{% if KOMA class
  \KOMAoptions{parskip=half}}
\makeatother

\usepackage{xcolor}
\usepackage[margin=2cm]{geometry}

\usepackage{longtable,booktabs,array}
\usepackage{calc} % for calculating minipage widths

\usepackage{etoolbox}
\makeatletter
\patchcmd\longtable{\par}{\if@noskipsec\mbox{}\fi\par}{}{}
\makeatother

\IfFileExists{footnotehyper.sty}{\usepackage{footnotehyper}}{\usepackage{footnote}}
\makesavenoteenv{longtable}

\usepackage{graphicx}
\makeatletter
\def\maxwidth{\ifdim\Gin@nat@width>\linewidth\linewidth\else\Gin@nat@width\fi}
\def\maxheight{\ifdim\Gin@nat@height>\textheight\textheight\else\Gin@nat@height\fi}
\makeatother

\setkeys{Gin}{width=\maxwidth,height=\maxheight,keepaspectratio}

\makeatletter
\def\fps@figure{htbp}
\makeatother


\newlength{\cslhangindent}
\newlength{\csllabelwidth}
\newlength{\cslentryspacingunit} % times entry-spacing
 {% don't indent paragraphs
  \setlength{\parindent}{0pt}
  \ifodd #1
  \let\oldpar\par
  \def\par{\hangindent=\cslhangindent\oldpar}
  \fi
  \setlength{\parskip}{#2\cslentryspacingunit}
 }%
 {}

\newcommand{\hyd}{$\text{H}_2$~}
\usepackage{float}
\usepackage{authblk}

\author[1]{Julen Larrucea}
\author[2]{Marita Oliv}
\author[2]{Jeanette Lorenz}
\affil[1]{Fraunhofer Institute for Manufacturing Technology and Advanced Materials IFAM, Bremen, Germany}
\affil[2]{Fraunhofer Institute for Cognitive Systems IKS, Munich, Germany}

\ifLuaTeX
  \usepackage{selnolig}  % disable illegal ligatures
\fi

\IfFileExists{bookmark.sty}{\usepackage{bookmark}}{\usepackage{hyperref}}
\usepackage{cleveref}
\IfFileExists{xurl.sty}{\usepackage{xurl}}{} % add URL line breaks if available
\hypersetup{
  pdftitle={Accuracy--Cost Trade-offs for Reference Electronic State Calculations on IBM Quantum Hardware: Session vs Single-Job Execution},
  pdfkeywords={quantum computing, variational quantum eigensolver, hydrogen molecule, IBM Quantum, benchmarking, NISQ},
  hidelinks,
  pdfcreator={LaTeX via pandoc}
}

\title{Accuracy-Cost Trade-offs for Reference VQE Calculations of H$_2$ on IBM Quantum Hardware}
\date{\today}

\begin{document}
\maketitle

\begin{abstract}

We present a hardware-validated reference dataset for variational
ground-state energy calculations of the hydrogen molecule H\(_2\) on
several IBM Quantum processors available in 2026. Using a standardized
workflow, we benchmark the impact of shot count, backend choice,
optimization strategy, and runtime variability on the achievable energy
accuracy relative to exact diagonalization. The resulting dataset and
analysis provide a transparent baseline for assessing the current
capabilities and limitations of IBM Quantum hardware for
quantum-chemistry applications, and are meant to ease the entry for new users by providing a comprehensive overview of choices and their effects as well as runtime efforts and costs that can be expected.

Across the configurations studied here, circuit simplification through tapered mappings provides the most consistent accuracy gains, resilience level 1 improves accuracy at a substantial cost premium, and session-based execution yields no systematic accuracy advantage over single-job execution despite markedly higher billed time.
\end{abstract}

\section{Introduction}
\label{sec:introduction}
Hybrid quantum-classical algorithms are currently the primary route for extracting physically meaningful quantities from noisy intermediate-scale quantum (NISQ) devices. 
Among these methods, the variational quantum eigensolver (VQE) has become a standard benchmark for molecular electronic-structure calculations on near-term hardware because it combines relatively shallow circuits with an iterative hybrid optimization loop \cite{Peruzzo2014,Kandala2017,McCaskey2019,Tilly2022}.

Despite its conceptual simplicity and widespread adoption, practical use of VQE on real quantum hardware remains challenging. In particular, users often lack reliable intuition for how common workflow parameters, such as shot count, backend choice, circuit mapping, or execution mode, translate into achievable accuracy, runtime, and billed cost.

This lack of intuition is especially pronounced for new or non-specialist users, for whom VQE often serves as an entry point into quantum chemistry on hardware. Accordingly, we focus deliberately on tutorial-aligned, out-of-the-box workflows: we use publicly available tooling and default workflows from Qiskit and Qiskit Nature with minimal user intervention, and we avoid bespoke circuit engineering or problem-specific noise tailoring. 
This design choice intentionally prioritizes representative user workflows over hand-optimized performance, so that the resulting trends are directly relevant to practitioners entering hardware-based quantum chemistry.

As a benchmark system, we consider the electronic ground state of the hydrogen molecule H\textsubscript{2}. As the smallest nontrivial molecular system, H\textsubscript{2} admits an exact classical solution at negligible cost and a compact qubit representation. It has therefore become a canonical test case for experimental VQE implementations \cite{Peruzzo2014,Kandala2017,Arute2020} and for studies of ansatz expressibility, optimization dynamics, and error mitigation techniques \cite{Sim2019,Sharma2020,Endo2021}. Because the underlying chemistry is fixed and simple, differences in performance across experiments can be attributed primarily to execution settings, backend characteristics, and workflow choices rather than to properties of the molecular problem itself.

Existing literature on VQE for small molecules largely emphasizes algorithmic developments, ansatz design, or simulator-based studies \cite{Wecker2015,Sim2019,Tilly2022}, while hardware demonstrations typically report results for fixed shot counts or single executions per configuration \cite{Kandala2017,McCaskey2019,Arute2020,Huggins2022}. Consequently, there is a lack of systematically collected, hardware-executed reference data that quantifies how standard, tutorial-aligned workflows behave across different backends, shot budgets, resilience settings, and execution modes.

The goal of this work is to provide such a reference to ease the entry into the execution of VQE on quantum processors and to set the expectations in terms of required resources. Using a standardized Qiskit Nature workflow for problem construction and ansatz generation, combined with a VQE optimization loop compatible with Qiskit~2.x primitives, we perform an extensive hardware study across multiple IBM Quantum backends, shot counts, circuit mappings, resilience levels, and both session-based and single-job execution modes. For each configuration, we record not only the deviation from the exact energy but also detailed execution metrics, including optimizer iterations, wall-clock time, quantum execution time, and billed time. The resulting dataset establishes a practical baseline for assessing accuracy-cost trade-offs in reference electronic-state calculations and provides empirically grounded guidance for users of current IBM Quantum hardware.

Section \ref{sec:methods} defines the benchmark problem, software stack, and execution methodology. Section \ref{sec:empirical_behavior} presents the empirical results for optimizer behavior, backend variability, shot count, mapper choice, resilience settings, and execution mode. Section \ref{sec:discussion} discusses the broader implications and limits of these observations, and Section \ref{sec:conclusions} concludes. Additional technical details and extended material are provided in the Appendix.
\section{Methods and Reference Workflow}
\label{sec:methods}

The reference workflow used throughout this study is designed to characterize the empirical behavior of standard hardware-executed VQE workflows under realistic user conditions.

Therefore, all calculations are performed using publicly available software components and tutorial-aligned defaults, with minimal user intervention.

All experiments are performed using the publicly available Qiskit software stack. Problem construction and ansatz generation rely on Qiskit Nature, while optimization routines are provided by \texttt{qiskit\_algorithms}. Hardware execution is carried out using the Qiskit Runtime service through the Estimator primitive, which provides a unified interface for expectation-value evaluation and access to execution metadata. All calculations are compatible with Qiskit~2.x. 

\begin{table}[ht]
\centering
\label{tab:software_versions}
\begin{tabular}{ll}
\hline
\textbf{Component} & \textbf{Version / Information} \\
\hline

Qiskit & 1.4.3 \\
Qiskit Nature & 0.7.2 \\
Qiskit Nature PySCF interface & 0.4.0 \\
Qiskit Algorithms & 0.3.1 \\
Qiskit Aer & 0.17.1 \\
Qiskit IBM Runtime & 0.40.1 \\
PySCF & 2.10.0 \\

\hline

IBM Quantum hardware access period &
August 2025 -- April 2026 \\

\hline
\end{tabular}
\caption{Software environment and hardware access period used in this study.
All calculations were performed using the versions listed here.}
\end{table}

\paragraph{Qiskit Nature workflow components.}

The molecular electronic structure problem was generated using
\texttt{PySCFDriver} and represented as an
\texttt{ElectronicStructureProblem} within Qiskit Nature.
Fermion-to-qubit transformations were performed using the
\texttt{JordanWignerMapper} and \texttt{ParityMapper}, including
a symmetry-tapered variant obtained via
\texttt{get\_tapered\_mapper}.
The variational ansatz was constructed using \texttt{UCC}
initialized with a \texttt{HartreeFock} reference state.

Exact reference energies were obtained using the
\texttt{NumPyEigensolver} as well as noisy and noiseless \texttt{AerSimulator} from \texttt{qiskit-aer} with the \texttt{FakeNairobiV2} backend for the noisy one.

\paragraph{Benchmark system and problem definition}

The benchmark system considered in this work is the electronic ground state of the hydrogen molecule H\textsubscript{2}. Problem construction follows the standard Qiskit Nature workflow. The molecular Hamiltonian is generated using the \texttt{PySCFDriver}, which computes the Hartree--Fock reference state and the second-quantized fermionic Hamiltonian via PySCF \cite{Sun2018}. The molecular geometry is fixed at an internuclear distance of $0.735$~\AA{} (i.e., H at $(0,\,0,\,0)$ and H at $(0,\,0,\,0.735)$ in \AA{}), with total charge zero and spin parameter zero, corresponding to a singlet state under the convention $\mathrm{spin} = 2S$.
This choice defines the minimal STO--3G active space with two spatial orbitals and two electrons and reproduces the introductory examples used in the Qiskit Nature documentation.

Although H$_2$ is chemically simple, its small size is an advantage here because it allows repeated hardware executions across many workflow settings while keeping the physical problem fixed.

The exact ground-state energy used as a reference throughout this work is obtained by direct numerical diagonalization of the qubit Hamiltonian using \texttt{NumPy}~\cite{harris2020array}. 
For the geometry and basis set considered, this yields a reference energy of
$E_\mathrm{ref} = -1.85727503\ \mathrm{a.u.}$
All reported VQE results are expressed as deviations from this exact value.
This ensures that all observed inaccuracies originate from the variational procedure, sampling noise, or hardware effects, and not from approximations in the Hamiltonian construction.

Using a fixed molecular problem and a fixed reference energy allows us to interpret all observed differences as workflow effects (ansatz size, backend behavior, sampling level, resilience, and execution mode), rather than changes in the physical model.

For the mapping from the fermionic Hamiltonian to the qubit representation, Jordan--Wigner and parity mapping are used with different levels of symmetry tapering, resulting in four representations. Qiskit Nature's built-in mapping and symmetry reduction utilities are used 
%\ab{Would be nice to know the exact methods used by qiskit nature here}\jl{check after the version table}
. Because all four ansatz choices are derived from the same Hamiltonian and differ only in mapping/tapering complexity, cross-configuration comparisons isolate implementation tradeoffs rather than problem-definition changes.
\begin{itemize}
\item \textbf{JW}: Jordan--Wigner mapping\cite{jordan1928uber} without tapering, resulting in a four-qubit observable.
\item \textbf{P}: Parity mapping\cite{seeley2012bravyi_kitaev} without tapering, also yielding a four-qubit observable.
\item \textbf{PF}: Parity mapping with particle-number tapering, reducing the observable to two qubits.
\item \textbf{PT}: Fully tapered parity mapping\cite{bravyi2017tapering}, exploiting all available symmetries and yielding a single-qubit observable.
\end{itemize}

\paragraph{Ansatz generation and circuit properties}

As variational ansatz, we use Qiskit’s unitary coupled-cluster ansatz (UCC)\cite{Peruzzo2014}, fully aligned with tutorial-level usage for the H$_2$ benchmark.

The Hartree--Fock state is used as the initial reference, with one repetition of the excitation operators, preserved spin symmetry, no generalized excitations, and real-valued parameters. This, together with the Hamiltonian representations fully specifies the four quantum circuits. Parameter vectors are initialized to zero for the optimization. 

To quantify the intrinsic complexity of each ansatz independently of backend-specific transpilation effects, we show in table~\ref{tab:ansatz_properties} for each mapping the number of qubits, circuit depth, total number of operations, and the number of two-qubit entangling gates of the circuits on the level of one- and two-qubit gates, i.e. obtained from \texttt{ansatz.decompose(reps=2)}. These metrics, which differ by more than an order of magnitude, capture the algorithmic cost of the circuit and provide a reference for interpreting hardware performance.
Their structural differences anticipate the later observation that mapper-dependent circuit simplification is the dominant predictor of empirical accuracy in this benchmark.

To make the resulting circuits concrete: the fully tapered PT ansatz
consists of a single parameterized rotation gate on one qubit, so that the entire VQE reduces to a one-dimensional optimization over one angle; the two-qubit PF circuit contains four \textsc{cx} gates and one variational parameter; and the four-qubit JW and P circuits
implement the full set of UCC excitation operators with 48 and 32
\textsc{cx} gates, respectively (see Table~\ref{tab:ansatz_properties}).
All four circuits are fully determined by the molecular specification, mapper choice, and UCC configuration stated above---following the standard ground-state workflow of the Qiskit Nature
documentation---and can be reproduced with the software versions
listed in Table~\ref{tab:software_versions}.

\begin{table}[t]
\centering
\caption{Structural properties of the ansatzes derived from the same H\textsubscript{2} Hamiltonian. All values are extracted from the decomposed circuits prior to transpilation and are independent of backend-specific gate sets.}
\label{tab:ansatz_properties}
\begin{tabular}{lcccc}
\toprule
Ansatz & Qubits & Depth & Total ops & CX gates \\
\midrule
JW & 4 & 96 & 186 & 48 \\
P  & 4 & 74 & 106 & 32 \\
PF & 2 & 17 & 23  & 4  \\
PT & 1 & 1  & 1   & 0  \\
\bottomrule
\end{tabular}
\end{table}

\paragraph{Execution modes}

For the execution on IBM quantum devices there are several modes that differ in how jobs are treated by the IBM Runtime scheduler. Here, we consider:

\begin{itemize}
\item \textbf{Session execution}: All iterations of a VQE run are executed within a single runtime session, during which the user retains exclusive access to the backend for a time window. This mode minimizes queuing delays which helps to limit calibration drift during iterative optimization.
\item \textbf{Single-job execution}: Each VQE iteration is submitted as an independent job. This mode is representative of execution on backends without session support or under access plans with limited runtime capabilities.
\end{itemize}

When we report timings, \textit{billed time} in session execution corresponds to the full session duration, whereas in single-job execution it is accumulated over all submitted jobs.

Throughout the paper, \emph{qtime} denotes the reported quantum execution time, while \emph{btime} denotes billed runtime. In single-job mode these quantities are effectively accumulated per-job execution charges, whereas in session mode billed time includes the full reserved session interval.

\paragraph{Shot definition and estimator precision}

In Qiskit~2.x the Estimator primitive does not accept an explicit shot count. Instead, sampling accuracy is controlled through a \texttt{precision} parameter that determines the target statistical uncertainty of expectation values. For interpretability and comparison with standard quantum chemistry
benchmarks, we label configurations by a \emph{nominal shot count} $N_{\text{shots}}$.
The term \emph{nominal shot count} is therefore used as a convenient label rather than as a literal runtime parameter.

The mapping between nominal shots and Estimator precision is defined as

\begin{equation}
\epsilon = \frac{1}{\sqrt{N_{\text{shots}}}} .
\end{equation}

\paragraph{Error mitigation and resilience levels}

The Qiskit Runtime Estimator provides a hierarchy of error-mitigation strategies through a single resilience level parameter \cite{IBMRuntime_Resilience,IBMRuntime_OptionsV2}. We benchmark resilience levels 0, 1, and 2. Level 0 applies no mitigation. Level 1 enables measurement-error mitigation based on calibration data.
Level 2 additionally enables stronger mitigation mechanisms such as zero-noise extrapolation and gate twirling under the runtime defaults used here.
%Level 2 additionally applies zero-noise extrapolation and gate twirling. 
All mitigation strategies are used with default parameters in order to reflect standard usage.

\paragraph{Circuit generation and transpilation variability.}
For each execution, circuits were generated and transpiled independently using the default compilation workflow provided by the software stack.
No attempt was made to reuse or fix a single compiled circuit across repeated runs.

This choice reflects a realistic user workflow, in which circuits are typically regenerated and recompiled when jobs are resubmitted, for example after modifying execution parameters or restarting an interactive session.
As a consequence, the reported results capture not only hardware noise and finite sampling effects, but also variability arising from stochastic elements of the transpilation and layout process.

For the small circuits studied here, the decomposition into
backend-native gates can be expected to vary little, as the gate
sequences are short and offer limited rewriting freedom. The mapping
of logical to physical qubits, however, has a larger impact: because
the circuits occupy at most four qubits on processors with over 100
physical qubits, the transpiler may assign a different subset of the
chip at each submission, and qubit-level variations in gate fidelities,
coherence times, and readout errors~\cite{krantz2019quantum} will shift the effective noise environment from run to run. This is a deliberate aspect of the study design: the goal is to characterize the accuracy and cost that a user following standard documentation would encounter without manual intervention, and the default transpiler behavior---including stochastic qubit placement---is part of that out-of-the-box experience. The reported run-to-run variability therefore represents the full spread a non-specialist user can expect, encompassing both temporal calibration drift and spatial qubit-quality heterogeneity.

\paragraph{VQE implementation and optimizers}

At the time of this study, the standard Qiskit Nature minimum-eigensolver workflow is not directly compatible with IBM Runtime execution using the V2 primitive interface. We therefore implement a custom VQE optimization loop built around the EstimatorV2 primitive, while keeping the remaining workflow components aligned with the standard Qiskit Nature pipeline.

Optimization is performed primarily using the Constrained Optimization BY Linear Approximations (COBYLA) algorithm~\cite{nocedal2006numerical}, which exhibits rapid and stable convergence for the low-dimensional parameter spaces considered here. For reference, we also include the Simultaneous Perturbation Stochastic Approximation (SPSA) optimizer~\cite{spall1992multivariate}. SPSA’s gradient approximation requires only two objective function evaluations per iteration regardless of parameter dimensionality, but its default configuration includes a calibration phase (e.g., tens of exploratory steps) before substantive optimization begins~\cite{spall1992multivariate, spsa_qiskit_docs}. In our preliminary tests, COBYLA converged to reasonable energy estimates within roughly 20 Estimator calls, whereas SPSA’s calibration and default step sequencing would cost at least three times as many total energy evaluations without hyperparameter tuning, rendering it disproportionately expensive for the problem sizes studied here.
The purpose of the SPSA comparison is therefore not to establish a general optimizer ranking, but to justify the use of COBYLA as the default optimizer for the remainder of this study.

\paragraph{Quantum backends} All experiments are executed on IBM Quantum devices. In particular, results are reported for the Heron~r1 QPU \texttt{ibm\_torino}, the Heron~r2 QPUs \texttt{ibm\_aachen}, \texttt{ibm\_fez}, and \texttt{ibm\_kingston}, the Heron~r3 \texttt{ibm\_pittsburgh}, the Eagle~r3 \texttt{ibm\_brussels}, and the (currently still in exploratory phase) Nighthawk~r1 \texttt{ibm\_miami}.

All hardware executions reported in this work were performed between August 2025 and April 2026, as indicated in Table~\ref{tab:software_versions}. Because device calibrations evolve continuously over time, results obtained at different dates may reflect both backend characteristics and temporal calibration variability.

\paragraph{Recorded metrics and dataset}

For each VQE execution, defined by a unique combination of backend, ansatz, nominal shot count, resilience level, optimizer, and execution mode, we record the final energy deviation from the exact solution, the number of optimizer iterations, wall time, quantum-processor execution time, billed time, and full per-iteration convergence histories. 

These recorded quantities support three complementary views used in later sections: accuracy (energy error), computational effort (iterations and time), and economic cost (billed usage).

Numerical accuracy is assessed using the absolute deviation of the measured ground-state energy from the fixed classical reference value obtained from diagonalizing the qubit Hamiltonian. We denote this metric as

\begin{equation}
  \label{eq:e_err_def}
  E_{err} = \left|E_{\mathrm{final}} - E_{\mathrm{ref}}\right|,
\end{equation}
where $E_{\mathrm{final}}$ is the final VQE energy and $E_{\mathrm{ref}}$ is the exact reference energy.
In several figures and tables we report the scaled quantity $E_{err}\times 10$ (a.u.) for readability.
Throughout the paper, smaller values of $E_{err}$ indicate higher accuracy relative to the exact ground-state energy.

Cost is reported using runtime-provided metrics, namely the quantum time and the billed time. For single job executions the quantum and billed time coincide and are the sum of the reported quantum usage time. For session executions the billed time includes the quantum time and the waiting time, in which classical computations are run but the quantum device is blocked.
% YES. Pretty sure that it's correct
% Check if correct: In both cases the quantum time includes time for classical pre- and post-processing, i.e. compling and scheduling, and is measured only in units of full seconds.
We do not report wall clock time for the whole process from submitting a job to receiving a result, as network latency times and queuing times depend too much on individual settings and submission time.

The analysis tracks three main dimensions: final energy accuracy, optimization effort, and execution cost, as functions of backend, mapper, nominal shot count, resilience level, optimizer, and execution mode.

\section{Empirical Behavior of Common Workflow Choices}
\label{sec:empirical_behavior}

This section presents empirical results intended as a practical reference for
the behavior of commonly used workflow choices in near-term quantum chemistry
calculations.
All results are obtained from executions on real quantum hardware and are
reported with an emphasis on observed accuracy, variability, and execution
cost under realistic conditions.

Due to the stochastic nature of quantum measurements and the presence of
hardware noise, all reported results reflect aggregated behavior over repeated executions.
Observed variability should therefore be interpreted as an inherent property
of current quantum hardware and runtime environments, rather than as numerical uncertainty arising from post-processing.

Each subsection focuses on a specific workflow option, namely the energy convergence, backend selection, measurement shot counts, circuit mappings,
error mitigation strategies and job execution mode. But certain choices are discussed with regards to different aspects in various sections, due to correlated effects.

Because several workflow choices are correlated in practice, some results are revisited from complementary perspectives across the following subsections.

\subsection{Optimizer and Classical Optimization Routine}
\label{subsec:optimizers}
The energy error depends on the accuracy of the subroutine to prepare a state and evaluate the expectation value of the Hamiltonian, and on the optimization process. VQE is not guaranteed to find the circuit parameters that correspond to the minimal energy, even in noiseless simulations \cite{oliv_2022, poggel_2025}. The incurred energy error of VQE thus needs to be seen as combination of statistical, finite measurement uncertainties, hardware noise effects and minimization deficiencies. To help disentangle these effects, we first discuss the convergence in the optimization routine and the choice of optimizer for the rest of the paper.

A detailed analysis of suitable optimizers can be found in \cite{mihalikova_best-practice_2022} and \cite{oliv_2022}. Here, we only exemplarily show a comparison of COBYLA and SPSA as accessible via Qiskit.

\begin{figure}[t]
    \centering
    \includegraphics[width=0.60\textwidth]{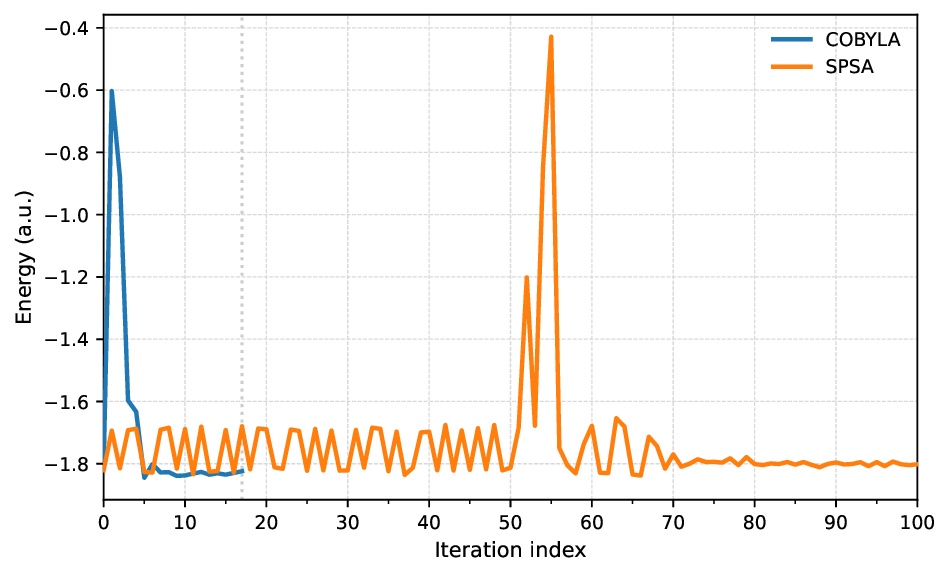}
    \caption{Convergence of COBYLA and SPSA energies on Kingston at 4096 shots. Only the first 100 SPSA evaluations are shown; beyond this point the SPSA energy oscillates around a narrow band without systematic improvement.}
    \label{fig:opt_convergence}
\end{figure}

Figure~\ref{fig:opt_convergence} shows the energy per iteration for an experiment on the Kingston processor at 4096 nominal shots using the PT ansatz and resilience level~0 for first 100 SPSA evaluations and the full COBYLA trace. Both runs start from identical initial parameters, namely the HF state, and use the same quantum configuration. The first 51 iterations of SPSA are objective function evaluations to calibrate the hyperparameters of the algorithm without circuit parameter optimization. After this, SPSA converges within 30 iterations with subsequent fluctuations around the value due to its stochastic update steps. COBYLA descends rapidly and reaches a stable energy after fewer than 15 iterations.
For all subsequent analyses, COBYLA is used as the default optimizer.

If an optimization converges at an energy level that does not correspond to the global minimum, this level usually is a local minimum induced by an excited state of the molecule. Hardware noise tends to bias expectation values towards higher energies, approaching the nuclear repulsion energy as a limiting case \cite{oliv_2022}. For the \hyd instance in this work, the first excited state lies at $-1.256$~a.u. Thus the convergence in figure~\ref{fig:opt_convergence} happens much closer to the ground state energy of $-1.857$~a.u. than to the excited state and the deviation from the exact ground state energy can be attributed to backend noise, rather than faulty convergence of the optimization routine.

\subsection{Backend Variability Under Fixed Workflows}
\label{subsec:backend_variability}

Before discussing algorithmic workflow hyperparameters, we show the backend-dependent behavior in order to impart the variability to be expected from using another device, as this is the parameter most difficult to reproduce. Later sections will revisit this variability for specific settings, while this section is meant to give a general idea of which deviations can be expected when executing on different backends.

\begin{figure}[htb]
\centering
\includegraphics[width=0.95\linewidth]{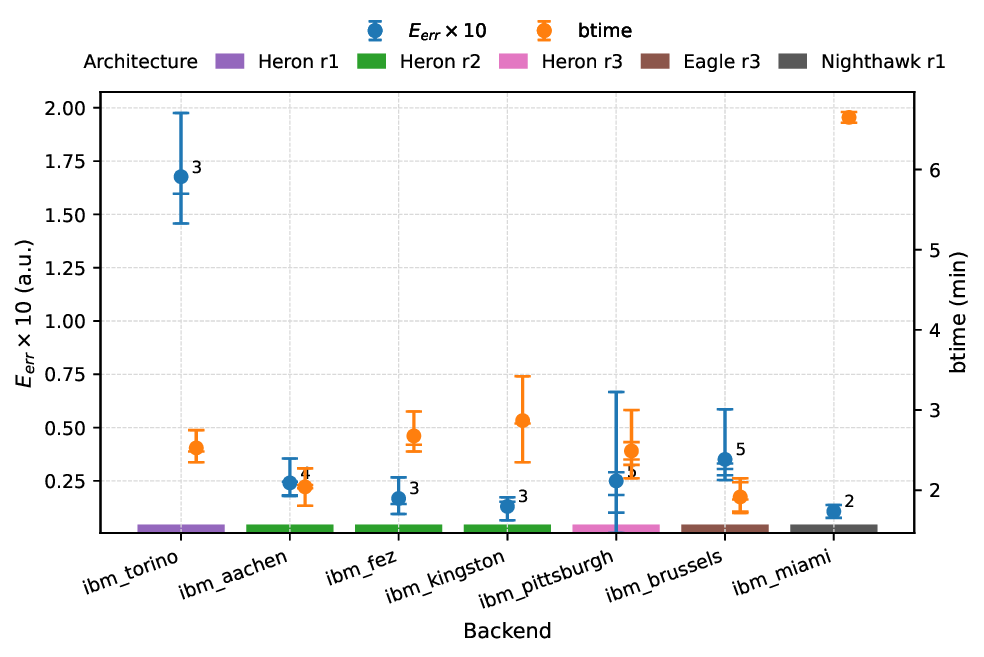}
\caption{Cross-backend comparison at PT, COBYLA, 1024 shots, resilience=0, session mode. $E_{err}=|E-E_{exact}|$ (reported as $E_{err}\times 10$ in a.u.); $btime$ denotes total billed time for one VQE execution. For each set the number of data points is denoted. %\ab{not clear to me what the numbers next to the blue dots are. If it relates to the number of runs, maybe should be done differently or be in the legend} \mo{description added to caption}
%and $qtime$ denotes quantum execution time.
}
\label{fig:02_backends_session}
\end{figure}

Figure~\ref{fig:02_backends_session} compares final energy errors and billed time across backends for a fixed workflow configuration (PT, COBYLA, 1024 nominal shots, resilience level 0, session mode). 
The spread within a particular backend can be attributed to changing properties over time and areas on the chip. Device properties and error probabilities considerably drift over time \cite{white_performance_2019} and repeated runs were collected over an extended period. Moreover, the properties between different qubits of a chip can differ substantially.
Device calibration data routinely show significant variability in coherence
times, readout errors, and gate fidelities across qubits on a single chip, sometimes spanning nearly an order of magnitude \cite{krantz2019quantum}. 

The variations between different backends reflect the device specific properties. 
The \texttt{ibm\_torino}, the first Heron~r1 processor which released 2023 \cite{blog_ibm_2023}, consistently delivers worse results than devices of later Heron builds, and than \texttt{ibm\_brussels}, an Eagle~r3 processor released 2024 \cite{trueman_ibm_2025}.

%Figure~\ref{fig:02_backends_session} summarizes the behavior for this configuration. The visual combines central tendency (mean), spread (min/max or full sample traces), and sample density annotations when available, enabling direct comparison across backends, mappers, resilience levels, and run modes.

\begin{figure}[htb]
\centering
\includegraphics[width=\linewidth]{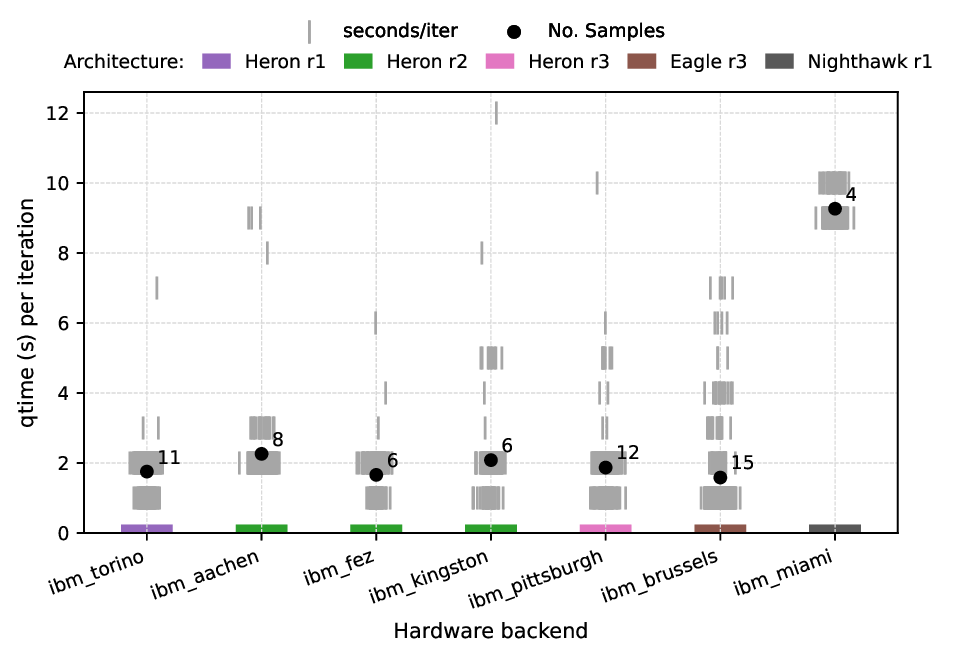}
\caption{Distribution of per-iteration quantum execution time (qtime) across backends. For each backend, the figure shows all individual \texttt{quantum\_seconds}, which are reported by IBM as integer values.
Values are shown as thin horizontal markers, together with a marker indicating the mean value. 
%\ab{the mismatch in colors between the base and the vertical markings is very confusing.} \jl{True! I need my home computer for that. I'll do it once I get home on friday (after sleeping).}
}
\label{fig:12_quantumseconds_per_iter_pt_1024_res0_mixed}
\end{figure}

%Figure~\ref{fig:12_quantumseconds_per_iter_pt_1024_res0_mixed} summarizes the behavior for this configuration. The visual combines central tendency (mean), spread (min/max or full sample traces), and sample density annotations when available, enabling direct comparison across backends, mappers, resilience levels, and run modes.

\begin{table}[htb]
\centering
\begin{tabular}{lrrrr}
\hline
Backend & rx & rz & x & sx \\
\hline
ibm\_miami      & 32.00 & 0.00  & 32.00 & 32.00 \\
ibm\_torino     & 32.24 & 54.69 & 32.24 & 32.24 \\
ibm\_fez        & 24.00 & 47.84 & 24.00 & 24.00 \\
ibm\_pittsburgh & 32.00 & 64.46 & 32.00 & 32.00 \\
ibm\_kingston   & 32.00 & 47.42 & 32.00 & 32.00 \\
ibm\_aachen     & 32.00 & 49.01 & 32.00 & 32.00 \\
ibm\_brussels   & --    & 0.00  & 60.00 & 60.00 \\
\hline
\end{tabular}
\caption{Single-qubit gate durations, averaged over qubits, at a single point in time (all values in nanoseconds). 
%\jl{Check this! Verify the "microseconds". The values 24–64 for single-qubit gates are in the range typical of nanoseconds for superconducting qubits, not microseconds. IBM Heron/Eagle single-qubit gates are typically 20–60 ns. If the values are indeed in nanoseconds, "microseconds" is an error by three orders of magnitude.}
}
\label{tab:gate_times}
\end{table}
The execution time in contrast is similar for most backends. 

Figure~\ref{fig:12_quantumseconds_per_iter_pt_1024_res0_mixed}  
reports the distribution of quantum execution time per individual circuit evaluation.
All data correspond to a fixed workflow configuration (PT mapping, 1024 shots, resilience level~0), aggregated over all optimization iterations and runs for each backend.
Notably, the vast majority of individual evaluations fall within a narrow
range of one to two seconds.

Table\ref{tab:gate_times} exemplarily lists the gate execution durations of several native single-qubit gates at a fixed point in time for different backends. The Rz rotation gate around the z-axis is a virtual gate for some backends and thus has no duration. The values are similar for all backends, explaining the similar per iteration circuit execution times. However, these gate durations also change over time with different calibrations.

Although per-evaluation time differs only modestly across backends, these small shifts accumulate over many objective-function evaluations and help explain the workflow-level timing differences observed earlier.
The results suggest that backend-dependent cost differences observed at the workflow level are not driven by sporadic long-running evaluations, but rather by systematic shifts in the typical per-evaluation execution time.
Accordingly, backend choice influences not only accuracy but also the
effective throughput of iterative algorithms under otherwise identical
settings.

\begin{figure}[htb]
\centering
\includegraphics[width=0.95\linewidth]{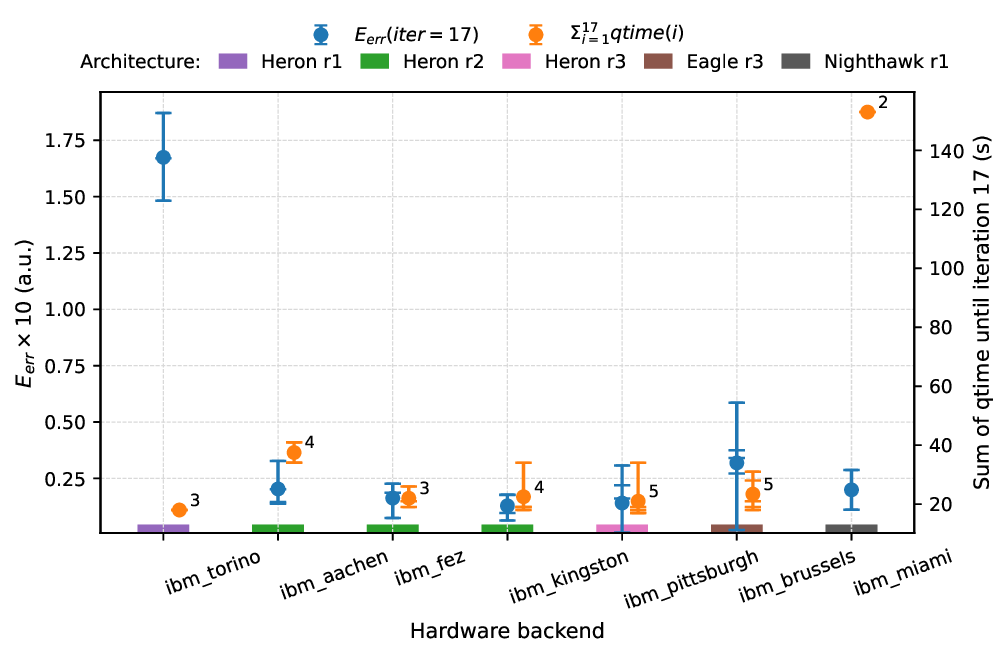}
\caption{Backend comparison at fixed optimization depth (iter=17) for error and cumulative qtime for same workflow parameters (PT, COBYLA, 1024 shots, resilience=0, session mode).}
\label{fig:11_backend_iter17_pt_1024_res0_mixed}
\end{figure}

To further disentangle backend-dependent effects from optimizer convergence behavior, Figure~\ref{fig:11_backend_iter17_pt_1024_res0_mixed} reports the energy deviation and accumulated quantum execution time at the same optimization step, namely iteration 17, for all backends, using an identical workflow configuration (PT mapping, 1024 shots, resilience level 0). 

Iteration 17 was selected because it is close to the point of convergence for most COBYLA runs, but little runs converge before.

A similar trend as in \cref{fig:02_backends_session} can be observed, indicating that the differences between backends do not stem from early or late convergence but backend-specific expectation value shift due to their noise properties.

%Figure~\ref{fig:11_backend_iter17_pt_1024_res0_mixed} summarizes the behavior for this configuration. The visual combines central tendency (mean), spread (min/max or full sample traces), and sample density annotations when available, enabling direct comparison across backends, mappers, resilience levels, and run modes.

Overall, the results show that backend choice can materially affect both accuracy and, for some cases, cost even when the circuit, optimizer, and nominal measurement budget are held fixed.

\subsection{Shot Count}
\label{subsec:shots_accuracy}

The number of measurement shots is one of the most visible control parameters in variational quantum algorithms and is often assumed to directly trade execution cost for improved accuracy. To assess this assumption empirically, here a systematic shot sweep is shown across multiple quantum backends while keeping all other workflow parameters fixed.

The shot counts considered span nearly four orders of magnitude.
In addition to a small number of exploratory low-shot executions (1, 16, and
64 shots), the main focus lies on the range from 256 to 8192 shots.
Among these, 1024 shots represents the most frequently sampled configuration
and serves as a natural reference point for comparison.
This distribution reflects common practical usage, where very low shot counts are primarily used for rapid testing, while moderate to high shot counts are employed for production runs.

\begin{figure}[htb]
\centering
\includegraphics[width=\linewidth]{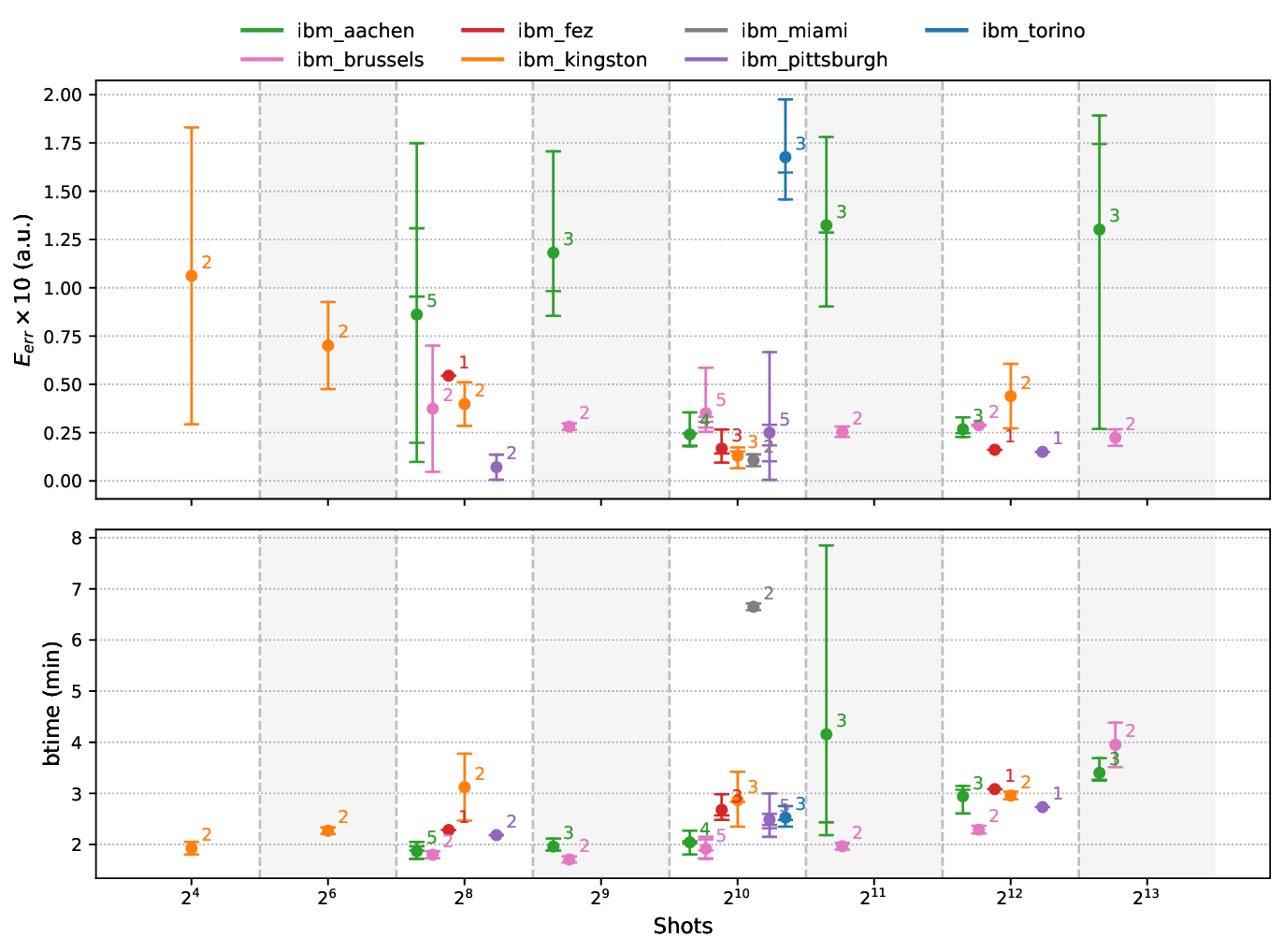}
\caption{
Energy error and billed time for different numbers of shots for multiple backends with PT mapper, resilience 0 in session mode. The dot marks the mean of the runs for which the individual results are shown as lines. The vertical bar spans the interval between the minimal and maximal value.
%Compact multi-backend shot-sweep view with per-sample markers and aggregated error bars.
}
\label{fig:13_shotsswipe_compact_all_backends_session}
\end{figure}

%Figure~\ref{fig:13_shotsswipe_compact_all_backends_session} summarizes the behavior for this configuration. The visual combines central tendency (mean), spread (min/max or full sample traces), and sample density annotations when available, enabling direct comparison across backends, mappers, resilience levels, and run modes.

Figure~\ref{fig:13_shotsswipe_compact_all_backends_session} provides a compact overview of the accuracy--billed time

trade-off associated with increasing measurement statistics across multiple IBM Quantum backends. In the appendix \ref{sec:supporting_info}, equivalent figures for individual backends are presented.
The expectation for an ideal device with no noise source other than stochastic sampling, assuming successful optimization, is that the standard deviation of the energy estimate decreases as $1/\sqrt{N_{\mathrm{shots}}}$ while the mean remains constant. Additional device noise shifts the mean of the distribution away from the exact value.
In practice, this shift depends on the backend and time of execution, due to drift.  

For most backends, little change in the energy error is seen for more than $2^8$ in this example. The quantum time, which directly corresponds to the billed time for the single job mode, on the other hand, rises linearly with the number shots with some backend specific variations.

On average across backends, the quantum time increases linearly with the number of shots, with a gradient of approximately $(1.08 \pm 0.19) \times 10^{-2}$~s\,/\,shot and an intercept of $27.0 \pm 5.5$~s (for figures see Appendix~\ref{sec:supporting_info}). Notably, the scaling for \texttt{ibm\_miami} deviates considerably from all other tested backends (compare Fig.~\cref{fig:13_shotsswipe_compact_all_backends_single_job}) because its circuit-layer operations per second (CLOPS) is an order of magnitude lower than that of the other backends. Its linear fit yields a gradient of $0.195 \pm 0.002$~s\,/\,shot and an intercept of $22.7 \pm 3.7$~s for the single set of available data points.

 To summarize, across all backends, increasing the number of shots generally leads to a reduction in the spread of observed energy deviations, while  most pronounced improvement is observed when moving from very low shot
counts to the intermediate regime around 256-1024 shots. Most backends exhibit a similar, linear scaling of the quantum time with the number of shots.
For the configurations sampled here, 1024 nominal shots emerges as a practical compromise between stability and cost.

\paragraph{Early-stage convergence structure across shots.}

To complement the fixed-iteration comparison in
Figure~\ref{fig:11_backend_iter17_pt_1024_res0_mixed}, we
analyze how quickly optimization progress is achieved in the first 5 and
10 iterations as a function of shot count.

To quantify this early-stage progress, we define a normalized convergence score.
For a run with energies $E_0,\ldots,E_T$, let
\begin{equation}
  \label{eq:convergence_score}
  E^\star = \min_{0 \le k \le T} E_k, \qquad
  C_n =
  \begin{cases}
  \displaystyle \frac{E_0 - \min_{0 \le k \le n} E_k}{E_0 - E^\star}, & \text{if } T \ge n \text{ and } E_0 \neq E^\star, \\[6pt]
  0, & \text{if } T \ge n \text{ and } E_0 = E^\star, \\[2pt]
  \mathrm{NaN}, & \text{if } T < n.
  \end{cases}
\end{equation}

In the following figures, \texttt{conv\_5} and \texttt{conv\_10} correspond to
$C_5$ and $C_{10}$ from Eq.~\eqref{eq:convergence_score}.

\begin{figure}[htbp]
\centering
\includegraphics[width=\linewidth]{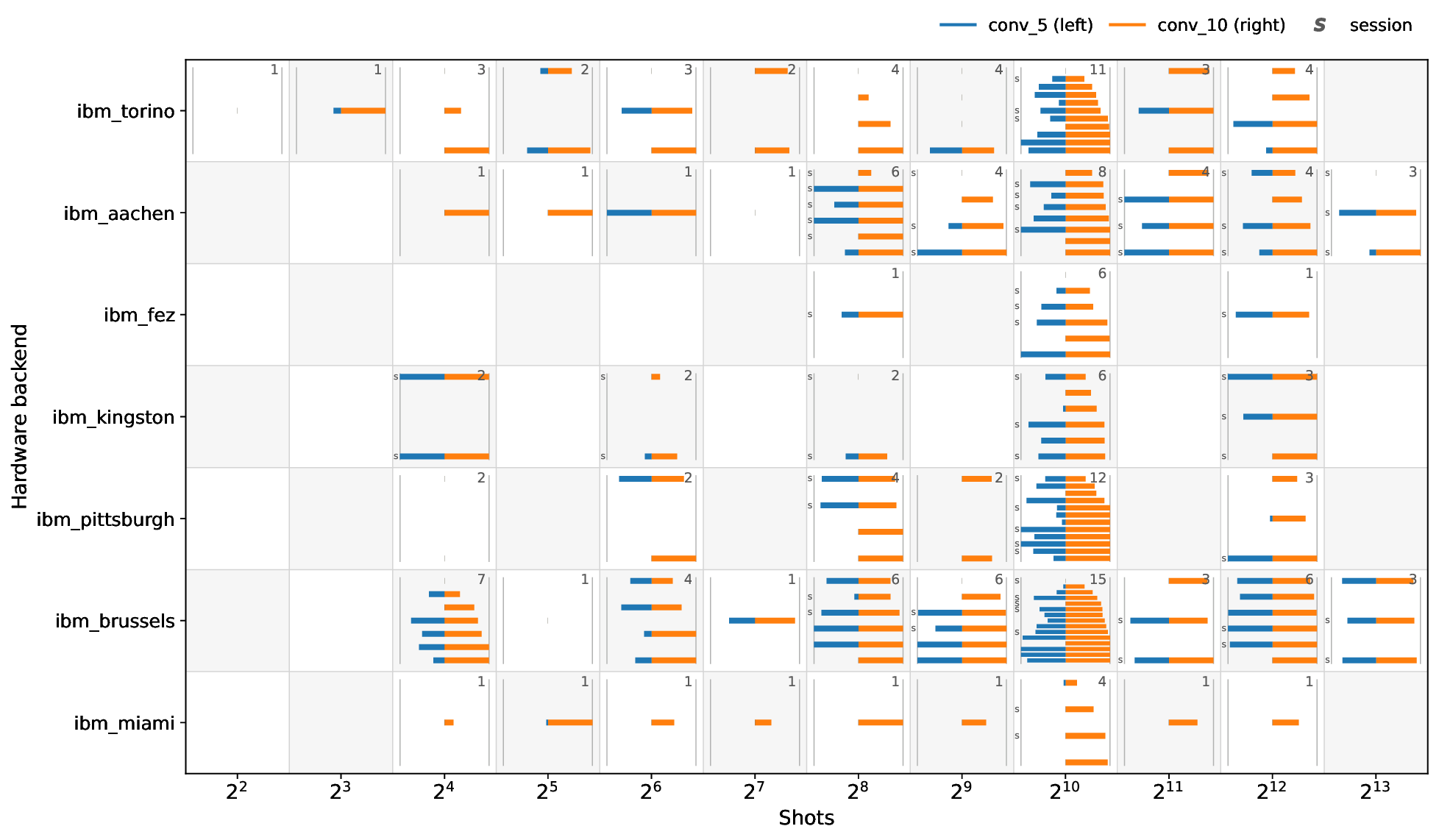}
\caption{Cell-wise early-convergence distributions ($conv_5$/$conv_{10}$) across backends and conditions. Bars show the energy after optimization iteration 5 or 10 relative to the energy at the end of the optimization process. A full bar means that the final energy is already reached in this iteration.}
\label{fig:16_convergencecells_res0_by_shots_pt}
\end{figure}

Figure~\ref{fig:16_convergencecells_res0_by_shots_pt} summarizes this behavior for
resilience level~0.
Thus, each cell compares the distributions of $C_5$ and $C_{10}$ under identical backend/shot conditions.

Each backend--shot cell contains paired per-sample bars (front:
\texttt{conv\_5}, back: \texttt{conv\_10}), so both spread and central
tendency can be inspected jointly.
The grid structure makes it clear that shot-dependent improvements are not
uniform across backends, and that increasing shots does not always produce a
proportional early-convergence gain.

%Figure~\ref{fig:16_convergencecells_res0_by_shots_pt} summarizes the behavior for this configuration. The visual combines central tendency (mean), spread (min/max or full sample traces), and sample density annotations when available, enabling direct comparison across backends, mappers, resilience levels, and run modes.

\subsection{Circuit Size and Mapping Overhead}
\label{subsec:mapper_overhead}

The choice of fermion-to-qubit mapping defines the problem given to the quantum computer and directly determines the size of the resulting quantum circuit. It therefore has a strong impact on both numerical accuracy and execution cost and should be discussed next.

The mappings considered include the Jordan-Wigner (JW), and Parity (P) both using four qubits,
Parity with two-qubit reduction (PF), and Parity with tapering to one qubit (PT) schemes.

\begin{figure}[htbp]
\centering
\includegraphics[width=0.49\linewidth]{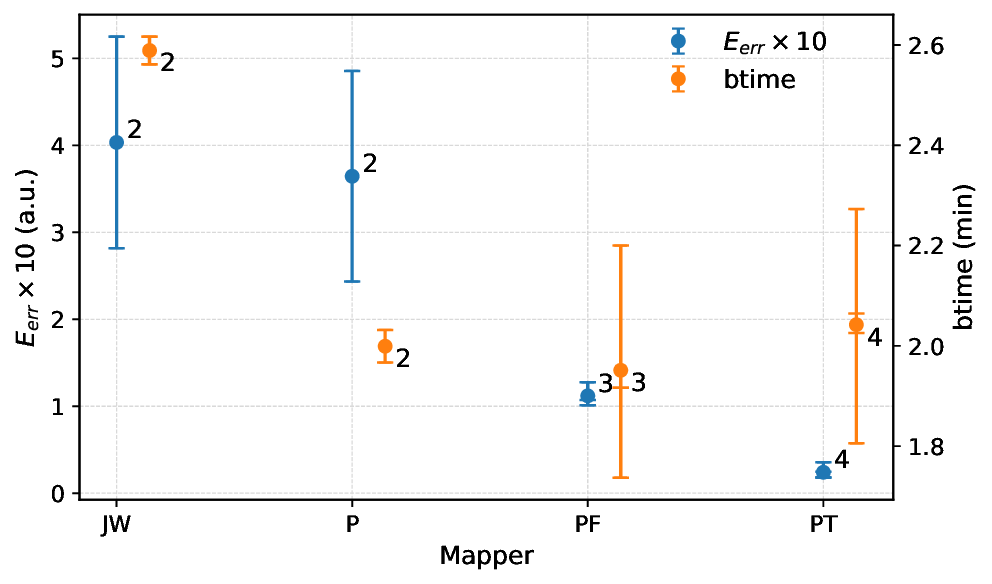}
\includegraphics[width=0.49\linewidth]{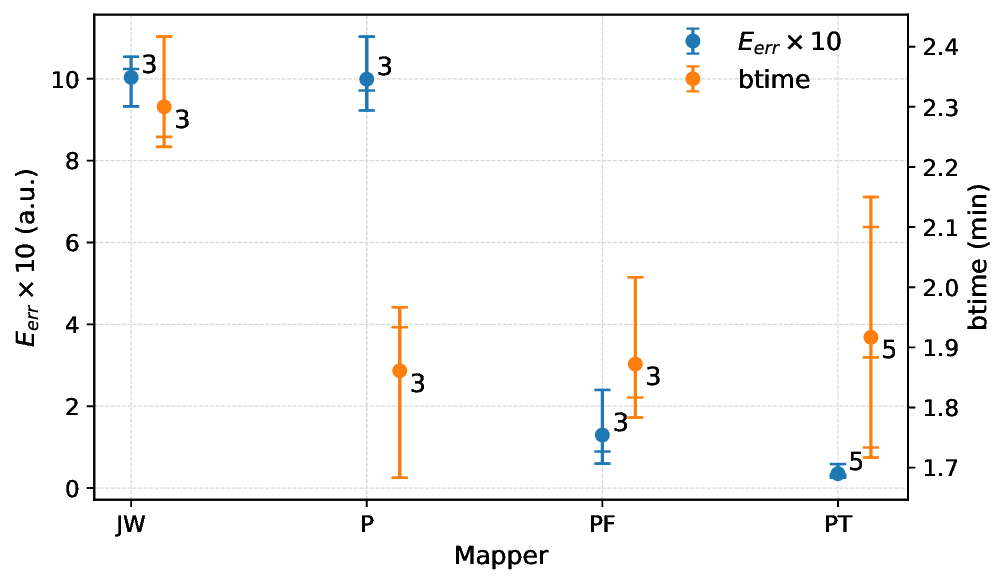}
\caption{Mapper comparison on \texttt{ibm\_aachen} (left) and ibm\_brussels (right) at COBYLA, 1024 shots, resilience=0, session mode. $E_{err}=|E-E_{exact}|$ (reported as $E_{err}\times 10$ in a.u.); $btime$ denotes billed time; bars span minimum to maximum value.}
\label{fig:04_mappers_session_ibm_aachen_brussels}
\end{figure}

Figure~\ref{fig:04_mappers_session_ibm_aachen_brussels}
displays the billed computing time and energy error for the different representations of the problem. Mappings that produce smaller and structurally simpler circuits consistently
yield lower energy deviations and reduced execution times. 

The energy error clearly correlates with the number of qubits in the circuit. The reduced depth and two-qubit gate count for the parity mapping compared to the Jordan-Wigner mapping does not lead to a clear improvement in the energy error, but to an evident reduction of the execution time. 

 The overall trend of lower errors for smaller circuits can be understood from error accumulation. With typical error rates of about $7\times10^{-3}$ for two-qubit gates and $3\times10^{-4}$ for single-qubit gates, we can roughly expect a gate error in about one out of three Jordan--Wigner (JW) circuits, whereas for the more compact particle-number-tapered parity (PF) mapping this probability is closer to one out of 27 circuits. In addition, qubit decoherence is proportional to the circuit execution time. Since readout errors are largely independent of circuit depth, they are expected to dominate for very small circuits (one or two qubits), while gate and decoherence errors increasingly contribute for deeper, larger circuits.

Table~\ref{tab:simulator_mapper_results} provides a noisy-simulator control using the same workflow under a fixed depolarizing noise model error rates ($p_1=0.003$, $p_2=0.02$). 

It shows the same mapper ordering observed on hardware (JW and P worst, PF intermediate, PT best), but at much smaller absolute error. This comparison is used only as an interpretive control: it confirms that mapper-dependent circuit complexity drives the trend direction, while the larger hardware magnitudes reflect additional real-device contributions.

\begin{table}[t]
\centering

\begin{tabular}{lrrrr}
\toprule
Mapper & Qubits & Energy (a.u.) & $|E-E_{\mathrm{ref}}|$ & Evals \\
\midrule
JW & 4 & -1.30257445 & 5.547e-01 & 24 \\
P & 4 & -1.41483325 & 4.424e-01 & 23 \\
PF & 2 & -1.78036555 & 7.691e-02 & 23 \\
PT & 1 & -1.85481687 & 2.458e-03 & 22 \\
\bottomrule
\end{tabular}
\caption{Aer-simulator VQE energies for H$_2$ (STO--3G) across mapper choices. Runs use nominal 16384 shots, COBYLA optimizer, 80 iterations for the optimization as maximum, resilience level 0, and noisy simulation.
}
\label{tab:simulator_mapper_results}
\end{table}

Among the workflow choices examined here, mapper-driven circuit simplification is the factor with the clearest and most consistent positive effect on final accuracy.

It is noteworthy that these effects persist despite the fact that all
considered quantum processing units provide a substantially larger number of
physical qubits than required by even the largest circuit studied here.
Since each execution effectively occupies the full device, execution cost is
not governed by qubit availability but by circuit depth, gate count, and
associated error accumulation.
The observed behavior therefore confirms that, under current hardware
conditions, circuit complexity is the dominant factor in determining both
accuracy and cost.

\begin{figure}[htb]
    \centering
    \includegraphics[width=\linewidth]{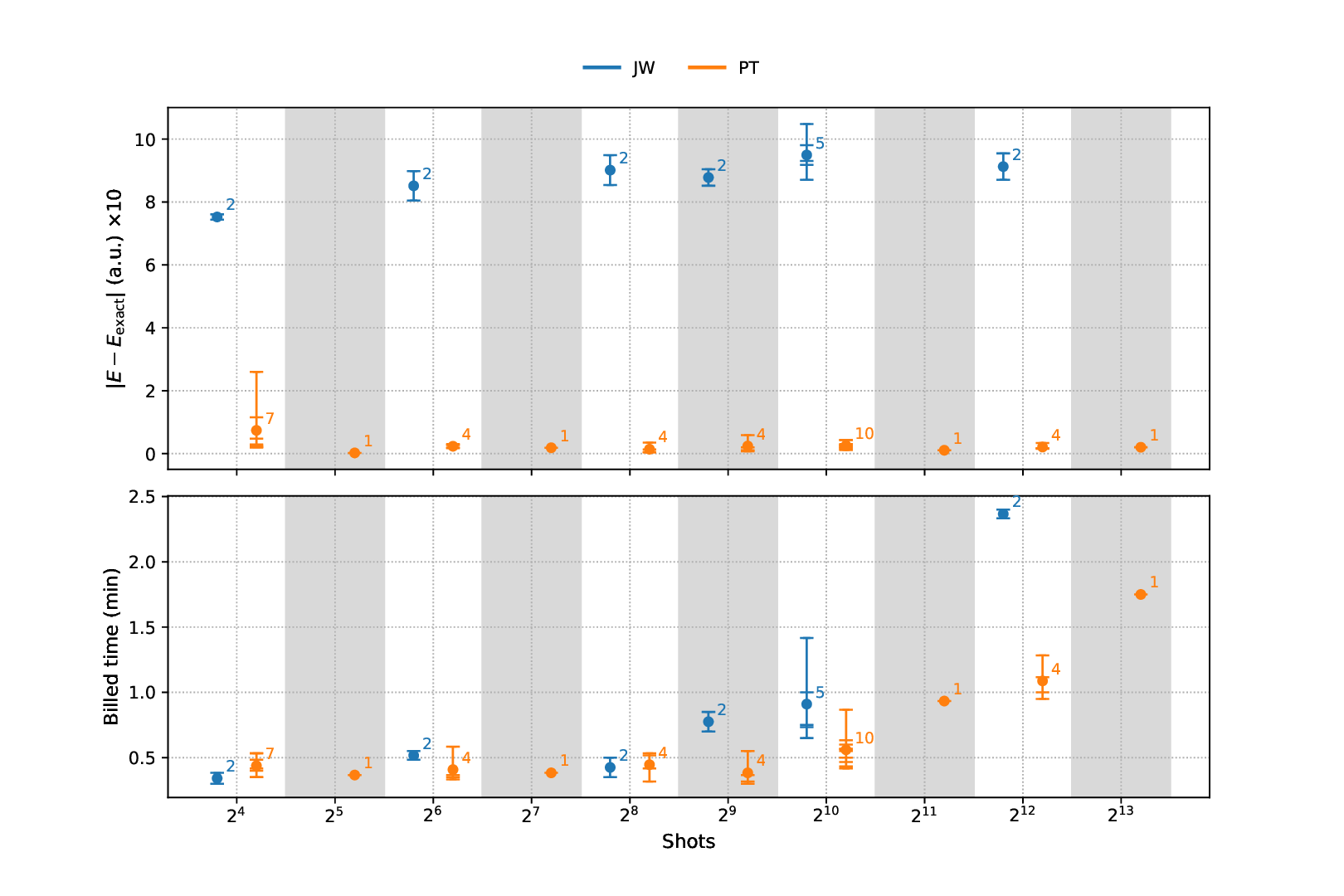}
    \caption{Energy error and billed time vs. number of shots for the JW and PT mapping (ibm\_brussels, resilience 0, job mode). Bars spread minimum to maximum value.}
    \label{fig:shotswipe_brussels_jw_pt}
\end{figure}
With this understanding of the effect of the choice of mapping, we revisit the choice of the number of shots. Figure~\ref{fig:shotswipe_brussels_jw_pt} shows the energy and billed time for different numbers of shots for the JW and PT mapper for ibm\_brussels. The energy shift between results from the two mappings is mostly independent of the number of shots. Similarly as for the PT circuits seen before, for JW  there is no jump in the energy due to a faulty optimization convergence to higher energy local minima for a relatively small number of shots. For larger problems one can not expect this to be the case as well, but that the required number of shots rises with the problem size as it is for variational algorithms for other problem classes  \cite{poggel_2025}. The billed time rises more steeply with the number of shots for the JW mapping than is has been seen for the PT mapper, as the execution time for a single execution of the deeper JW circuit is longer than for PT. 

\subsection{Resilience and Error Mitigation Effects}
\label{subsec:resilience_effects}

Error mitigation strategies are commonly employed to improve the accuracy of
near-term quantum computations by post-processing measurement results.
In this subsection, the empirical impact of readout error mitigation and zero noise extrapolation, i.e. resilience levels 1 and 2 in the Qiskit framework, is evaluated relative to the unmitigated baseline (resilience level 0), using
identical circuits, shot counts, and execution settings.

The relative improvement in energy accuracy obtained with resilience levels 1 and 2, compared to resilience level 0, is summarized in \cref{fig:06_resimprovement_session} and the corresponding tables can be found in appendix \ref{sec:supporting_info}. 

\begin{figure}[htbp]
\centering
\includegraphics[width=0.49\linewidth]{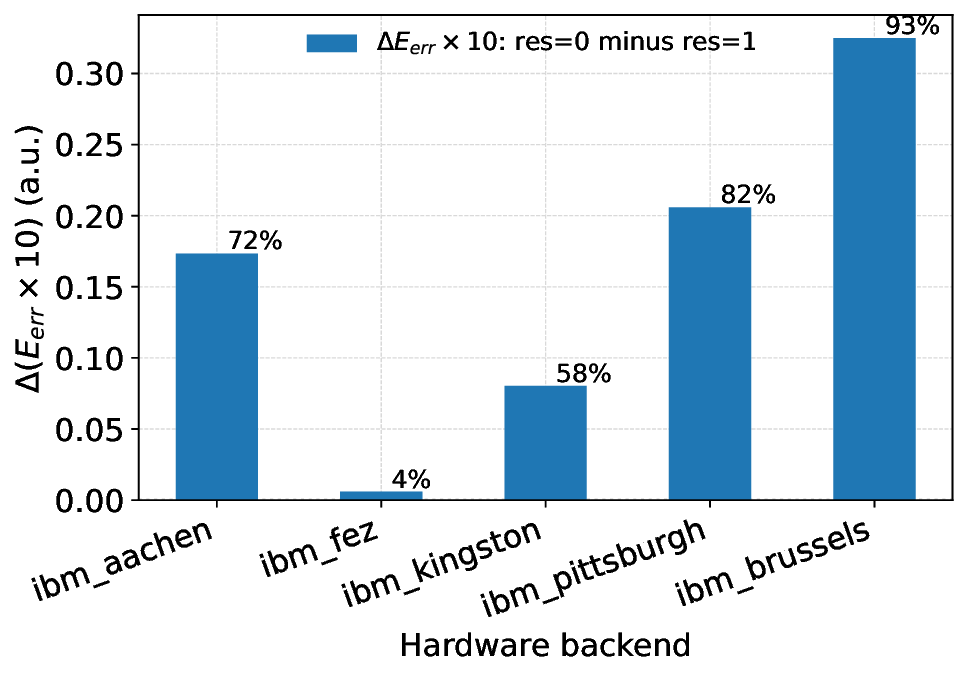}
\includegraphics[width=0.49\linewidth]{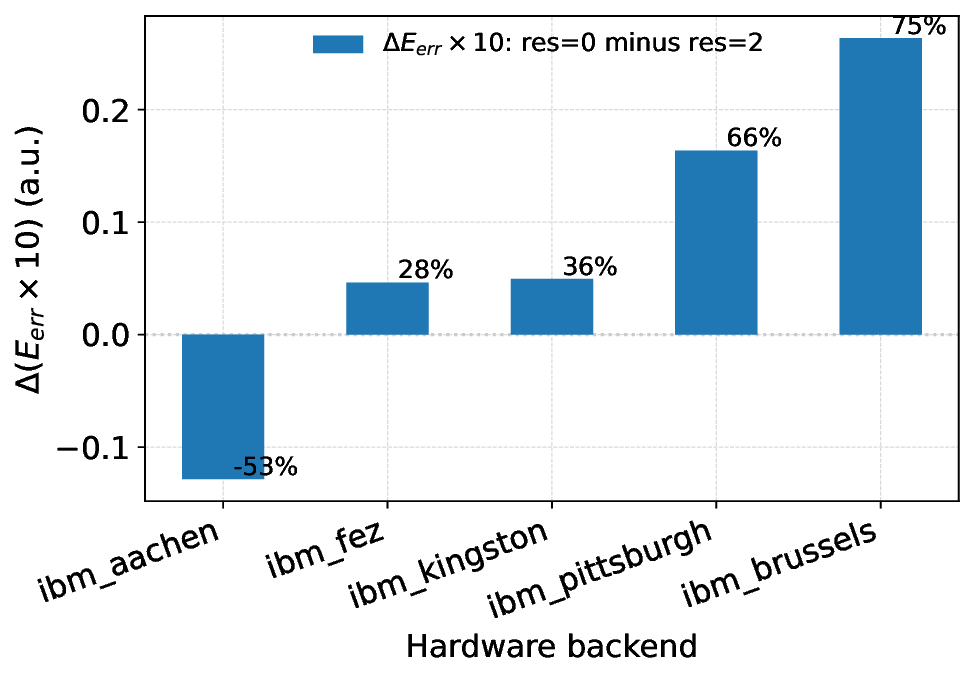}
\caption{Energy-improvement for resilience level 1 (left) and level 2 (right) relative to the energy obtained with level 0 (no mitigation) across backends.}
\label{fig:06_resimprovement_session}
\end{figure}

Within this dataset, resilience level 1 improves the mean result for all examined backends, although the evidence is stronger for some backends than for others because repeat counts are uneven. 
The accuracy improvement remains positive for resilience level 2 in all cases except \texttt{ibm\_aachen}, where resilience level 2 results in a degradation of the final energy estimate relative to resilience level 0.
This observation highlights that increased mitigation strength does not
guarantee improved results.
In particular, higher resilience levels may amplify statistical noise or
introduce additional bias when sampling is limited, leading to diminished or
even negative returns.

\begin{figure}[htbp]
\centering
\includegraphics[width=\linewidth]{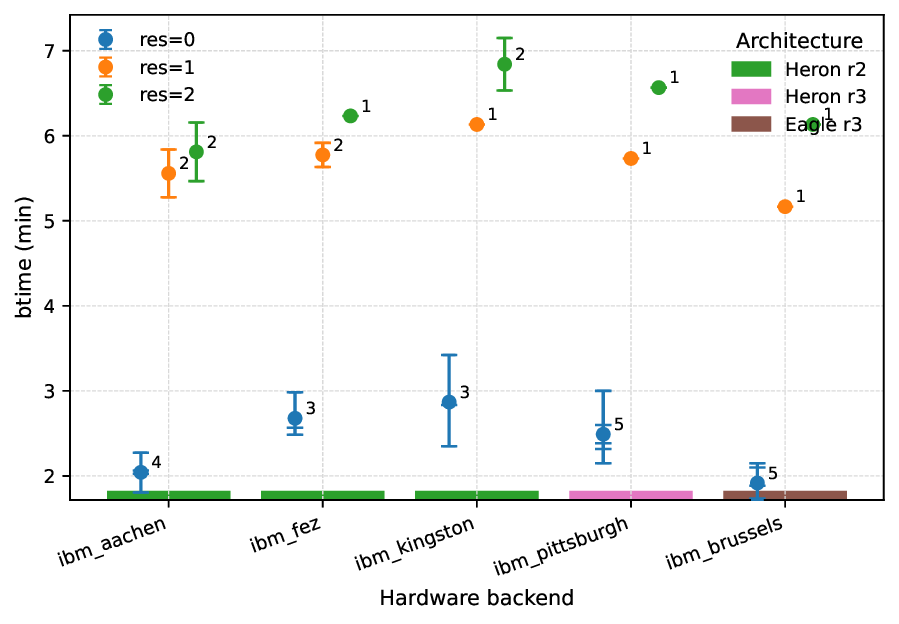}
\caption{Resilience impact on $btime$ at PT, COBYLA, 1024 shots, session mode. $E_{err}=|E-E_{exact}|$ (reported as $E_{err}\times 10$ in a.u.); $btime$ denotes billed time and $qtime$ denotes quantum execution time.}
\label{fig:03_resilience_time_session}
\end{figure}

The execution-time overhead associated with resilience levels 1 and 2 is
summarized in
Figure~\ref{fig:03_resilience_time_session}.

%Figure~\ref{fig:03_resilience_time_session} summarizes the behavior for this configuration. The visual combines central tendency (mean), spread (min/max or full sample traces), and sample density annotations when available, enabling direct comparison across backends, mappers, resilience levels, and run modes.

As expected, higher resilience levels incur additional execution cost due to
increased classical post-processing and additional circuit
evaluations.
The magnitude of this overhead varies across backends but is systematic,
with resilience level 2 consistently being more expensive than both the
unmitigated baseline and resilience level 1, as expected.

Thus, error mitigation can enhance accuracy across different backends, but its effectiveness is variable. This indicates that mitigation should be applied selectively, considering the trade-off between accuracy gains and the costs and performance implications. For practical applications, the decision to implement higher mitigation levels should be based on whether the accuracy improvements justify the additional time and resources required, compared to the accuracy achieved with lower mitigation.

% Maybe we could rename this as "Execution Mode Overview"
% or put the next 3 subsections as subsubsections of this one?

\subsection{Session vs Single-Job Execution}
\label{subsec:session_vs_single_job}
 The IBM Quantum platform allows quantum workloads to be executed either as \emph{single, standalone jobs} or grouped into \emph{runtime sessions}. The latter is thought to be beneficial in particular for variational algorithms. For the VQE execution, the classical optimization is not done on the IBM cloud, but the result from the QPU is passed back to the local, classical computer after each execution. QPU calculations of consecutive iterations cannot be parallelized. Using session mode avoids entering the queue for the job to be scheduled every single time.  This is meant to reduce the overall wall clock time from submission of the first job to retrieving the last result. A short time frame limits the hardware property drift that occur during the execution. As the optimizer thus samples from a more consistent objective function, this could improve the output quality of the variational algorithm. 

\begin{figure}[htbp]
\centering
\includegraphics[width=\linewidth]{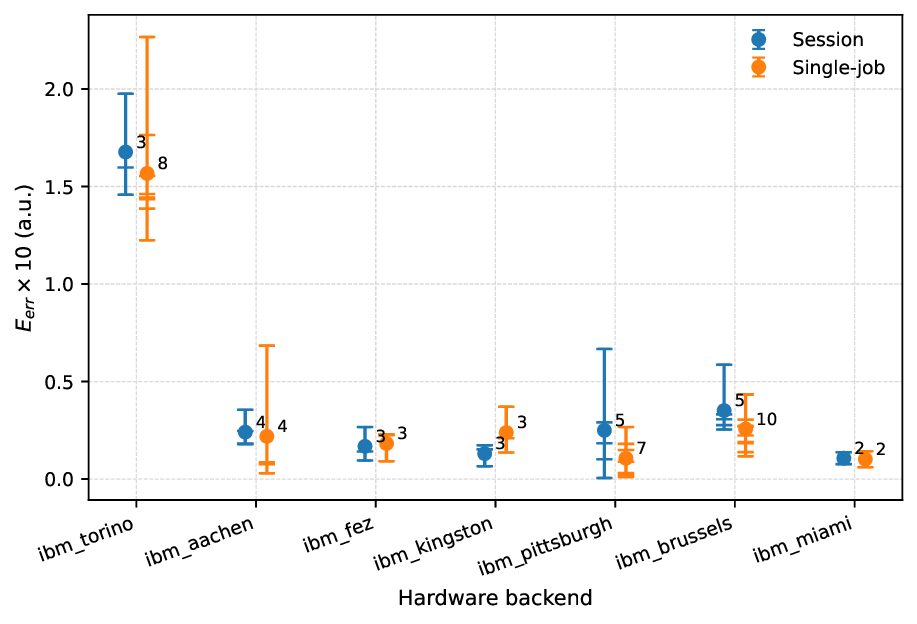}
\caption{Session vs single-job comparison of $E_{err}$ across backends for workflow with 1024 shots and PT mapper without error mitigation. $E_{err}=|E-E_{exact}|$ (reported as $E_{err}\times 10$ in a.u.).%; $btime$ denotes billed time and $qtime$ denotes quantum execution time.
}
\label{fig:08_energy_session_vs_single_1024}
\end{figure}

Figure~\ref{fig:08_energy_session_vs_single_1024} compares the energy error from runs in session and job mode for different backends. The results are consistent with each other and no improvement is seen for session mode. This correlates with the short overall time taken for the execution.
The mean time between two consecutive execution of jobs, including latency, parameter optimization step, further pre- and post-processing on user side and queuing times have been between 7~s and 5~m~19~s depending on the backend for job mode. In tendency, the first few iterations take longer than later ones. For session mode, there are initial, relative long waiting times of several minutes, followed by mean time between executions of 10~s to 3~m~32~s, depending on the backend. In total, the mean time between executions have been below 5~m~:36~s for all backends. For both modes, most executions have happened within tens of seconds of each other. 
The overall wall clock execution time of the whole VQE algorithm for the configuration shown in \cref{fig:08_energy_session_vs_single_1024} has been on average around 32~m for session mode and around 44~m for job mode.  
These numbers are highly dependent on several external factors like details of the contract for access to the devices and the current workload from other users.  They are not meant as reference, but to explain the little energy differences seen. In particular, we expect little drift over the duration of few hundred to thousand single-qubit gate circuits \cite{rudinger_probing_2019}.

%Figure~\ref{fig:08_energy_session_vs_single_1024} summarizes the behavior for this configuration. The visual combines central tendency (mean), spread (min/max or full sample traces), and sample density annotations when available, enabling direct comparison across backends, mappers, resilience levels, and run modes.

\begin{figure}[htb]
\centering
\includegraphics[width=\linewidth]{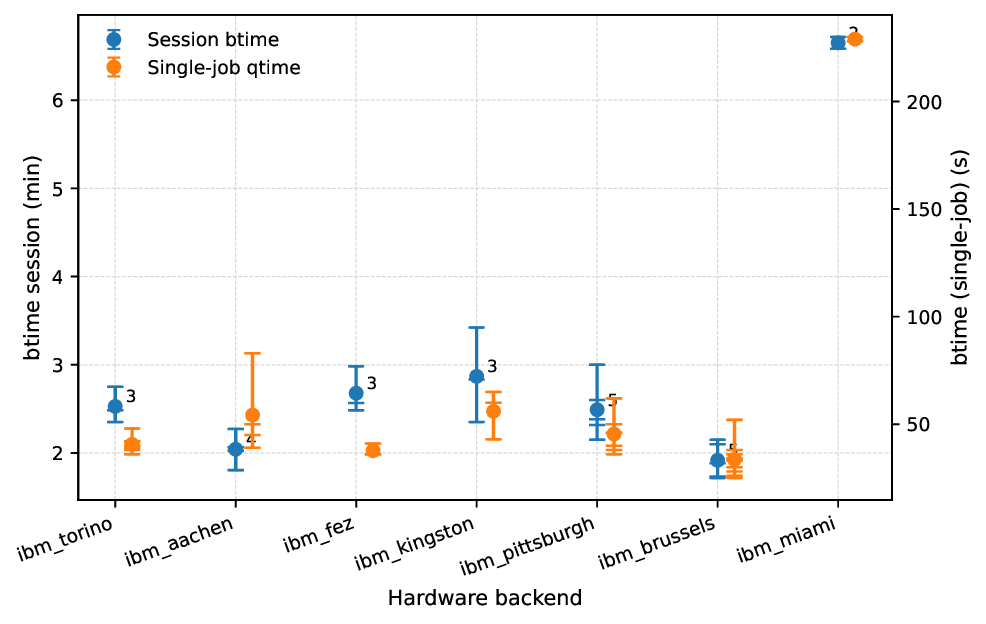}
\caption{Session vs single-job timing comparison across backends at matched algorithmic settings (PT, 1024 shots, resilience level 0).
%\ab{the double x axis is very confusing here. Two plots are necessary or using only btime for both, since btime = qtime for single-jobs right? } % \mo{Is the specification of settings correct?; right axis label needs adaption (sum qtime(sec)); talk about billed time in both cases}
}
\label{fig:07_time_session_vs_single_qsum_1024}
\end{figure}

Figure~\ref{fig:07_time_session_vs_single_qsum_1024} compares billed time for session executions (minutes) against summed quantum time in seconds, which corresponds to the billed time, for single-job runs across backends at fixed reference settings.
The backend ranking is broadly consistent between the two measures: backends that are faster in single-job mode also tend to be faster in session mode.
However, the absolute scales differ dramatically: session execution is billed on the order of minutes, whereas single-job execution is billed on the order of seconds. 

%Figure~\ref{fig:07_time_session_vs_single_qsum_1024} summarizes the behavior for this configuration. The visual combines central tendency (mean), spread (min/max or full sample traces), and sample density annotations when available, enabling direct comparison across backends, mappers, resilience levels, and run modes.

\section{Discussion}
\label{sec:discussion}

%The results presented in Sections~3 and~4 allow for a critical assessment of several commonly suggested best practices in current quantum algorithm workflows. While many of these practices are well motivated conceptually, their practical impact depends strongly on problem size, hardware characteristics, and cost models. In this section, we synthesize the main lessons learned and discuss their implications for typical users of near-term quantum hardware.

The results in Section \ref{sec:empirical_behavior} allow us to test several commonly recommended workflow choices against hardware-executed evidence for a fixed reference problem. For this dataset, increasing nominal shots beyond an intermediate regime yields diminishing accuracy returns (Section \ref{subsec:shots_accuracy}), mapper-driven circuit simplification provides the largest consistent accuracy gains (Section~\ref{subsec:mapper_overhead}), resilience level 1 improves accuracy robustly while resilience level 2 is more variable and more expensive (Section \ref{subsec:resilience_effects}), and session execution does not provide a systematic accuracy benefit but incurs substantially higher billed time than single-job execution (Sections \ref{subsec:session_vs_single_job}).

\paragraph{Revisiting common best practices}

Several workflow choices are often recommended as broadly beneficial, including the use of runtime sessions, or increased shot counts. The empirical evidence presented here suggests that these recommendations should be applied with greater caution.

In particular, session-based execution is frequently advertised as a performance optimization for iterative algorithms. Our results show that, for small- to medium-scale VQE workloads, sessions do not provide a systematic accuracy advantage and introduce a substantial cost overhead. 
%This finding contradicts the implicit assumption that sessions are universally beneficial and highlights the importance of validating such recommendations empirically for the workload of interest.

Similarly, although more complex circuits can lead to better results for noiseless devices, they do not necessarily translate to improved results on current hardware. In several cases, simpler configurations achieve comparable or better accuracy at significantly lower cost \cite{boutakka_benchmarking_2025, Kandala2017}.
%increased circuit complexity, whether through mapper choice or deeper ans\"atze, does not necessarily translate into improved results on current hardware. In several cases, simpler configurations achieve comparable or better accuracy at significantly lower cost \cite{Kandala2017}.
%\mo{do you have a reference where this is implied? otherwise I would remove it -> I meant it the other way round. I reformulated to make more clear why one could potentially expect better results for more complex circuits}

\paragraph{Lessons for non-expert users} 

A central motivation for this study was to provide guidance for users who are not specialists in quantum hardware or algorithm engineering. From this perspective, several practical lessons emerge.

First, the default workflow does not automatically reach good results, even for minimal problem sizes. Minimal tuning often is not sufficient to achieve precision within chemical accuracy, i.e., 1~kcal/mol ($\approx 1.6$~mHa). Instead, the default workflow can act as a template for more more optimized
solutions.

Second, many advanced features increase complexity and cost without delivering commensurate benefits, particularly when applied indiscriminately. In particular, for short circuits readout noise is the dominant source of errors. Thus, resilience level 1 can be a good choice to mitigate those, while higher resilience levels increase the cost but bring no considerable accuracy improvements.

Third, the use of session mode should be considered carefully, especially if using a pay-as-you-go plan. For overall execution times in the order of minutes to few hours, the accuracy might be affected little while the billed times increases from the order of seconds to minutes for subroutines. We thus suggest to first run an experiment in job mode and if necessary and reasonable switch to session mode later.

The observations suggest that non-expert users should prioritize simplicity and transparency over aggressive optimization. Incremental changes—tested one at a time—are more effective than adopting multiple advanced features simultaneously without clear performance diagnostics.

\paragraph{Limits of default workflows}

While the results favor simple workflows for the present benchmark, this conclusion should not be interpreted as a blanket endorsement of default settings. Rather, it delineates the regime in which defaults remain effective.

As problem sizes grow, circuits deepen, and optimizers require more iterations, the balance between execution overhead and algorithmic runtime can shift. In such regimes, more coordinated execution modes, stronger error mitigation, or customized scheduling strategies may become necessary to maintain performance and accuracy. However, the
present results indicate that, for the small to medium-scale workloads examined here, this transition occurs at larger problem sizes than is often assumed in practical workflows \cite{cerezo2021variational,temme2017error}.
%\mo{reference? where is it usually assumed?}

The key limitation of default workflows, therefore, is not correctness but scalability. Recognizing when a workflow leaves the regime explored in this study is essential for making informed methodological choices.

A further limitation is that repeat counts are not fully balanced across all configurations, particularly for some resilience and shot settings, so the dataset is best interpreted as a broad empirical benchmark rather than a perfectly uniform factorial study.

The present conclusions should therefore be read as statements about the benchmark regime studied here: shallow to moderate VQE circuits for a minimal molecular problem under contemporary IBM Runtime workflows.

\paragraph{Generalizability beyond the \hyd benchmark}
%\texorpdfstring{H$_2$}{H2}

The H$_2$ molecule serves as a minimal yet nontrivial test case, enabling extensive statistical sampling and controlled comparisons across backends and workflow choices. While this simplicity is a strength for methodological analysis, it necessarily limits direct extrapolation to larger molecular systems.

Nevertheless, several conclusions are expected to generalize qualitatively. In particular, the absence of systematic accuracy gains from session execution and the strong dominance of cost considerations for small problem instances are likely to persist for other small molecules and toy models commonly used in benchmarking studies.

Future work will be required to identify the precise problem sizes and circuit complexities at which the observed trends break down. Until then, the present results provide a grounded reference point for understanding the practical behavior of common quantum computing workflows on current hardware.

\section{Conclusions}
\label{sec:conclusions}

In this work, we have presented a systematic empirical study of common workflow choices in variational quantum chemistry calculations on current cloud-based quantum hardware. By analyzing accuracy, variability, and cost across execution modes, backends, and circuit constructions, we provide practical guidance grounded in quantitative evidence rather than heuristic recommendations.

\subsection{Key takeaways for practical workflows}

For the H$_2$ reference workflows examined in this study, several
practical conclusions can be drawn.

First, accuracy is strongly influenced by circuit complexity.
Mapper choices that reduce circuit size, particularly the PF and
PT mappings, consistently produce smaller energy deviations across
the tested backends.

Second, increasing the nominal shot count beyond a moderate range
provides diminishing improvements in accuracy for the configurations
examined here.

Third, resilience level~1 provides a clear and consistent improvement
in energy accuracy across the tested backends. However, this improvement
comes at the expense of a significant increase in computational cost,
as the additional error mitigation procedures require extra circuit
executions and classical post-processing. Resilience level~2 introduces
further overhead and does not consistently produce additional accuracy
gains for the configurations studied.

Finally, the comparison between session and single-job execution shows
no systematic difference in the achieved energy accuracy. This behavior
is observed both for the PT mapping, which produces the smallest circuits
in this study, and for the JW mapping, which results in larger circuits.
In both cases the resulting energies are statistically comparable between
the two execution modes. However, the billed time associated with session
execution is substantially larger than the accumulated quantum execution
time required for single-job execution. Under the conditions examined
here, single-job execution therefore provides a substantially more
cost-efficient approach for this class of workloads.

Based on these findings, we suggest that practitioners adopt single-job execution as the default strategy for exploratory studies, benchmarking, and routine quantum chemistry calculations on present-day hardware. Session execution should be introduced selectively and only after its benefits have been demonstrated empirically for the specific workload under consideration.

\subsection{Outlook}

As quantum hardware, runtime environments, and billing models continue to evolve, the balance between execution strategies is likely to change. Larger systems, deeper circuits, and more demanding optimizers may eventually shift workflows into regimes where session-based execution or more sophisticated scheduling becomes advantageous.

The methodology and reference data presented here provide a baseline against which such future developments can be assessed. By making workflow choices explicit and measurable, this work aims to support informed, transparent, and cost-effective use of near-term quantum computing resources in quantum chemistry and beyond.

A natural next step is to identify the problem size, circuit complexity, and optimization depth at which the present recommendations cease to hold.

\hypertarget{acknowledgements}{%
\section{Acknowledgements}\label{acknowledgements}}

The authors gratefully acknowledge support from the
\textbf{Fraunhofer internal program ``Quantum Now''}
(Editions 5 and 6), which provided both quantum computing
resources and personnel funding essential for the development
and execution of this work.

This work was carried out within the project
``Evaluierung von Quantencomputerressourcen für elektronische
Strukturberechnungen (EQeS), Teilvorhaben: Chemie und Analyse
auf Anwendungsebene'', funded by the German Federal Ministry
of Research, Technology and Space (Bundesministerium für
Forschung, Technologie und Raumfahrt, BMFTR) under contract
number 13N17338 as part of the Research Program Quantum Systems
(Forschungsprogramm Quantensysteme).

The research is part of the Munich Quantum Valley, which is supported by the Bavarian state government with funds from the Hightech Agenda Bayern Plus.

This research also made use of IBM Quantum services. The views
expressed are those of the authors and do not necessarily reflect
the views or positions of IBM or the IBM Quantum team.

The authors gratefully acknowledge the associate partners
Algorithmiq and the Donostia International Physics Center (DIPC),
and especially Dr.~Stephan Knecht, for helpful discussions and support.

\bibliographystyle{IEEEtran}
\bibliography{bibliography}

\clearpage
\appendix
\section{Supporting Information}
\label{sec:supporting_info}

\subsection{VQE Workflow: Detailed Object-Level Overview}
\label{subsec:si_vqe_workflow}

This subsection expands the end-to-end VQE chain used in this paper, from
molecular input to hardware-measured energy and timing outputs. The same
pipeline is reused in all experiments, while shots, resilience settings,
mapper choice, backend, and execution mode (session/single-job) are varied.
For a broader overview of VQE workflow steps \cite{Tilly2022} provides a well structured review.

\paragraph{Step 1: Driver (chemistry model instantiation).}
The workflow starts with \texttt{PySCFDriver}, which receives the molecular
specification (atomic geometry, basis set, charge, and spin). It executes the
classical electronic-structure setup and returns an
\texttt{ElectronicStructureProblem}. At this stage, no quantum circuit exists
yet: this object defines the physics target to be translated into qubit form.

\paragraph{Step 2: Electronic structure problem and second quantization.}
From the problem object, the fermionic Hamiltonian in second quantization is
obtained as
\begin{equation}
\label{eq:si_h_fermionic}
\hat{H}_{\mathrm{el}}
=
\sum_{pq} h_{pq}\, a_p^\dagger a_q
+ \frac{1}{2}\sum_{pqrs} h_{pqrs}\, a_p^\dagger a_q^\dagger a_r a_s,
\end{equation}
where $a^\dagger, a$ are fermionic creation/annihilation operators and
$h_{pq}, h_{pqrs}$ are one- and two-electron integrals in the selected basis.
This is the operator defining the target ground-state energy.

\paragraph{Step 3: Mapper (fermionic operator $\rightarrow$ qubit operator).}
The fermionic operator is transformed via mapper
$\mathcal{M}\in\{\mathrm{JW},\mathrm{P},\mathrm{PF},\mathrm{PT}\}$:
\begin{equation}
\label{eq:si_h_qubit}
\hat{H}_{q} = \mathcal{M}\!\left(\hat{H}_{\mathrm{el}}\right).
\end{equation}
Different mappers can change the resulting Pauli decomposition and effective
qubit count (including possible symmetry-based reductions/tapering). This is a
major source of the mapper-dependent behavior analyzed in the main text.

\paragraph{Step 4: Ansatz construction (state family).}
Given $\hat{H}_q$, a parameterized trial-state family is prepared using a
Hartree--Fock reference and UCC-style excitations:
\begin{equation}
\label{eq:si_ansatz}
\lvert \psi(\boldsymbol{\theta}) \rangle = U(\boldsymbol{\theta})\lvert \phi_0\rangle.
\end{equation}
There has been a plethora of other ansatz types suggested as well. The ansatz determines the reachable subspace and circuit complexity; together
with mapper and transpilation, this affects both accuracy and runtime.

\paragraph{Step 5: Estimator + classical optimizer loop.}
For each parameter vector $\boldsymbol{\theta}$ proposed by the optimizer, the
Estimator evaluates
\begin{equation}
\label{eq:si_vqe_obj}
E(\boldsymbol{\theta})=
\langle\psi(\boldsymbol{\theta})\rvert \hat{H}_q \lvert\psi(\boldsymbol{\theta})\rangle.
\end{equation}
The optimizer updates $\boldsymbol{\theta}$ iteratively until its stopping
criterion is met, producing a trajectory
$E_0,E_1,\ldots,E_T$ and final value $E_{\mathrm{final}}$. This trajectory is
the basis for the convergence analyses included in the figures and tables.

\paragraph{Output metrics used in this article.}
The main accuracy indicator is
\begin{equation}
\label{eq:si_eerr}
E_{err}\times 10
=
\left|E_{\mathrm{final}}-E_{\mathrm{exact}}\right|\times 10 \quad \text{(a.u.)}.
\end{equation}
Runtime is tracked with two complementary quantities:
\begin{itemize}
\item \textbf{qtime}: quantum execution time (\texttt{quantum\_seconds}),
reported per iteration/job and used for accumulated cost-style comparisons.
\item \textbf{btime}: billed runtime associated with the execution context
(especially relevant for session-mode analyses).
\end{itemize}

\paragraph{Why this object-level view matters for interpretation.}
The reported trends are not caused by a single factor: they emerge from the
interaction of driver/problem definition, mapper-induced operator structure,
ansatz expressibility, estimator noise (shots, resilience), backend properties,
and optimizer dynamics. 
%Figure~\ref{fig:si_vqe_workflow} is included to keep this dependency chain explicit when interpreting mapper, backend, and session/single-job comparisons.

\subsection{Additional Data}
\label{subsec:si_additional_artifacts}

This subsection includes tables that list the values shown graphically in the figures in the main text and additional figures.

\begin{figure}[htb]
\centering
\includegraphics[width=0.7\linewidth]{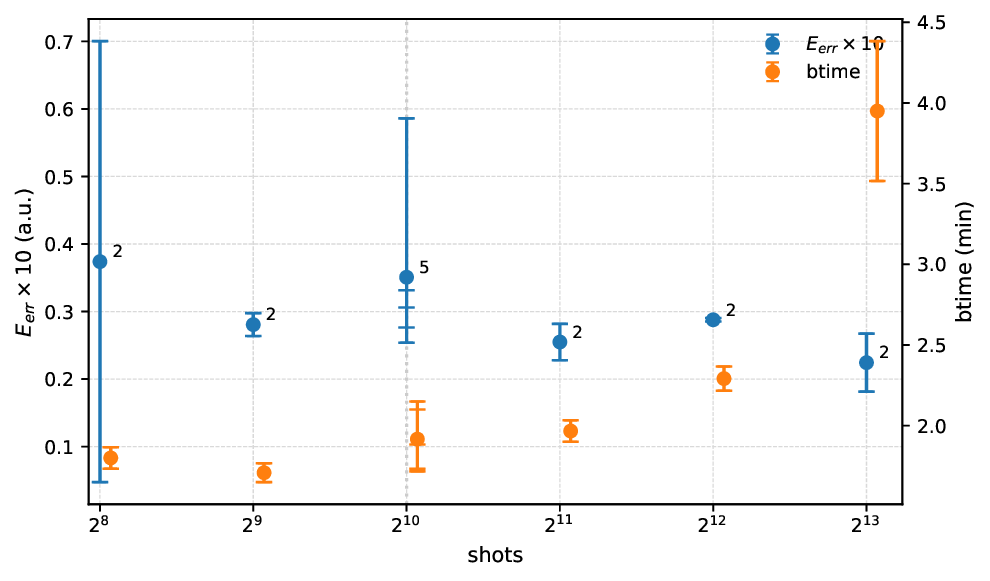}
\caption{Shot sweep on \texttt{ibm\_brussels} for PT, COBYLA, resilience=0, session mode. $E_{err}=|E-E_{exact}|$ (reported as $E_{err}\times 10$ in a.u.); $btime$ denotes billed time and the vertical bars spread minimum to maximum value.}
\label{fig:01_shotsswipe_session_ibm_brussels}
\end{figure}

\begin{figure}[htb]
\centering
\includegraphics[width=0.74\linewidth]{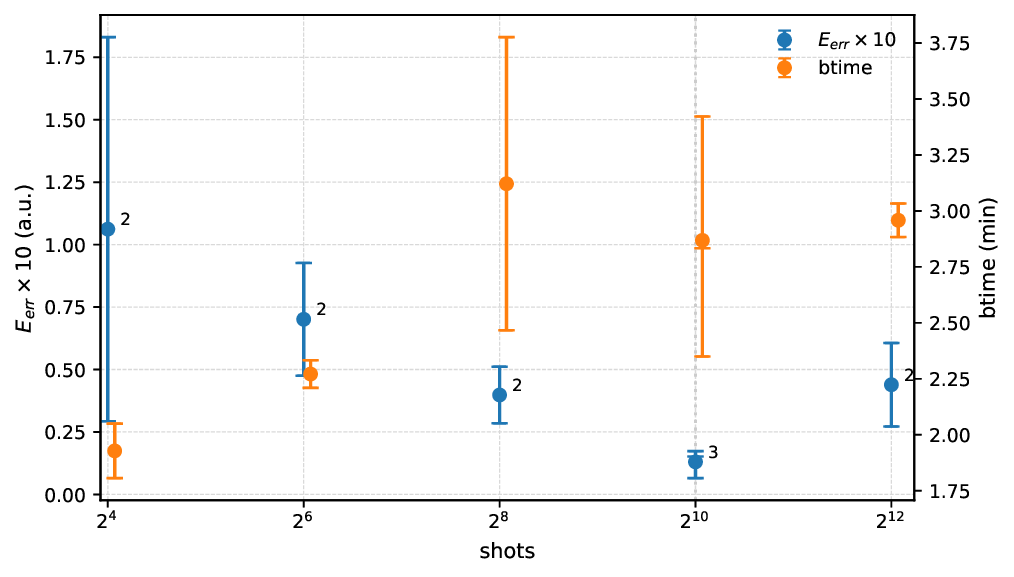}
\caption{Shot sweep on \texttt{ibm\_kingston} for PT, COBYLA, resilience=0, session mode. $E_{err}=|E-E_{exact}|$ (reported as $E_{err}\times 10$ in a.u.); $btime$ denotes billed time and the vertical bars spread minimum to maximum value.}
\label{fig:01_shotsswipe_session_ibm_kingston}
\end{figure}

\paragraph{Number of shots}
Figures \ref{fig:01_shotsswipe_session_ibm_brussels} and \ref{fig:01_shotsswipe_session_ibm_kingston} show the energy error and billed time for different numbers of shots as seen in \cref{fig:13_shotsswipe_compact_all_backends_session} separately for the backends ibm\_brussels and ibm\_kingston.

%Figure~\ref{fig:01_shotsswipe_session_ibm_brussels} summarizes the behavior for this configuration. The visual combines central tendency (mean), spread (min/max or full sample traces), and sample density annotations when available, enabling direct comparison across backends, mappers, resilience levels, and run modes.

%Figure~\ref{fig:01_shotsswipe_session_ibm_kingston} summarizes the behavior for this configuration. The visual combines central tendency (mean), spread (min/max or full sample traces), and sample density annotations when available, enabling direct comparison across backends, mappers, resilience levels, and run modes.

\begin{figure}[htb]
\centering
\includegraphics[width=\linewidth]{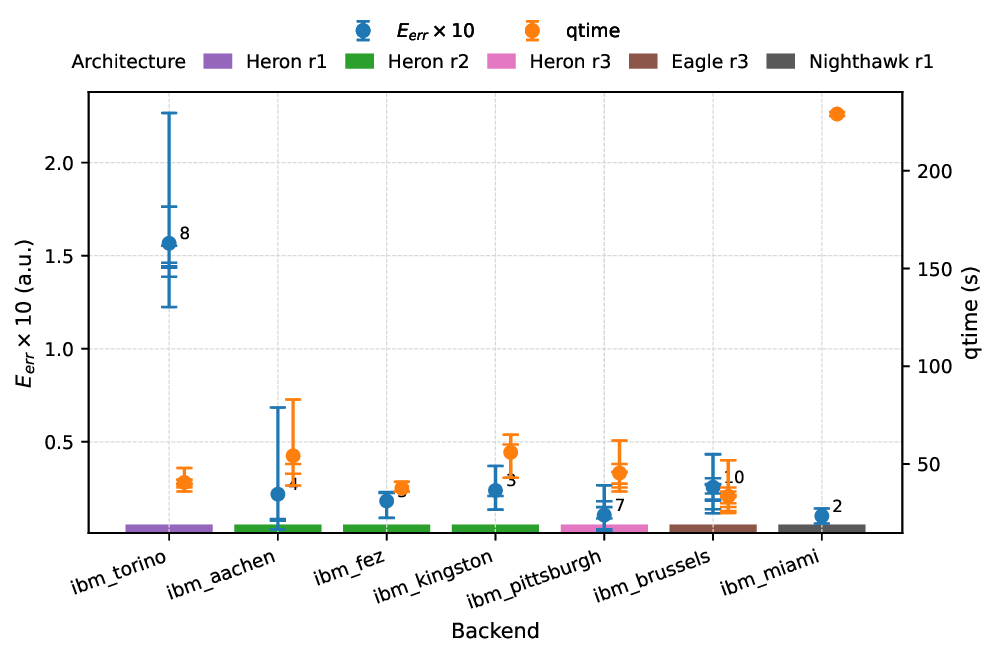}
\caption{Cross-backend comparison at PT, COBYLA, 1024 shots, resilience=0, single-job mode. $E_{err}=|E-E_{exact}|$ (reported as $E_{err}\times 10$ in a.u.); $btime$ denotes billed time and $qtime$ denotes quantum execution time.}
%\mo{unit on label right y axis missing}}
\label{fig:02_backends_single_job}
\end{figure}

Figure~\ref{fig:02_backends_single_job} summarizes the behavior for this configuration. The visual combines central tendency (mean), spread (min/max or full sample traces), and sample density annotations when available, enabling direct comparison across backends, mappers, resilience levels, and run modes.

\begin{figure}[htb]
\centering
\includegraphics[width=0.8\linewidth]{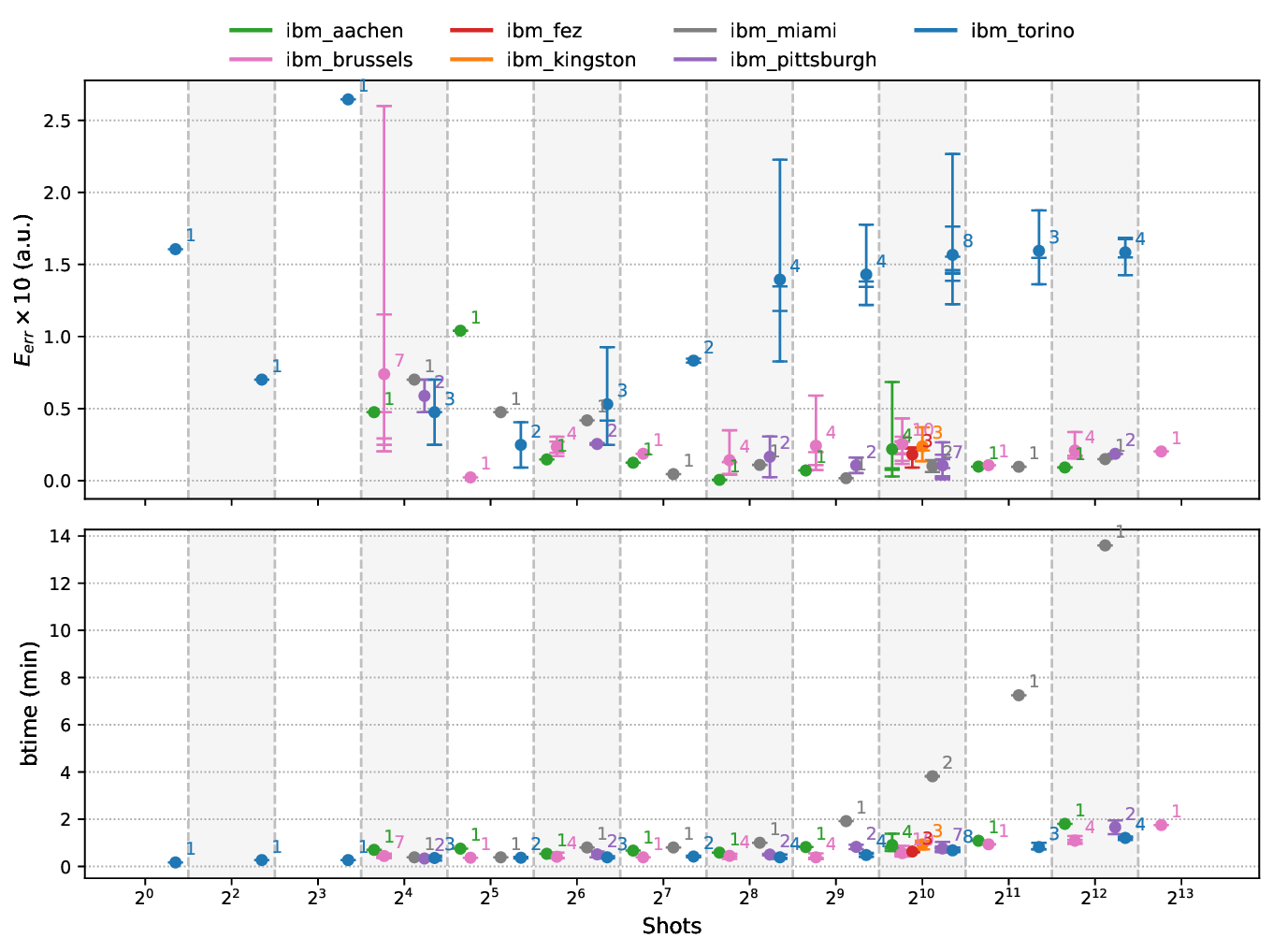}
\caption{Compact multi-backend shot-sweep view with per-sample markers and aggregated error bars, that span minimum to maximum value, for PT mapper, resilience level 0 in single job mode. }
\label{fig:13_shotsswipe_compact_all_backends_single_job}
\end{figure}

%Figure~\ref{fig:13_shotsswipe_compact_all_backends_single_job} summarizes the behavior for this configuration. The visual combines central tendency (mean), spread (min/max or full sample traces), and sample density annotations when available, enabling direct comparison across backends, mappers, resilience levels, and run modes.

\begin{figure}
    \centering
    \includegraphics[width=0.49\linewidth]{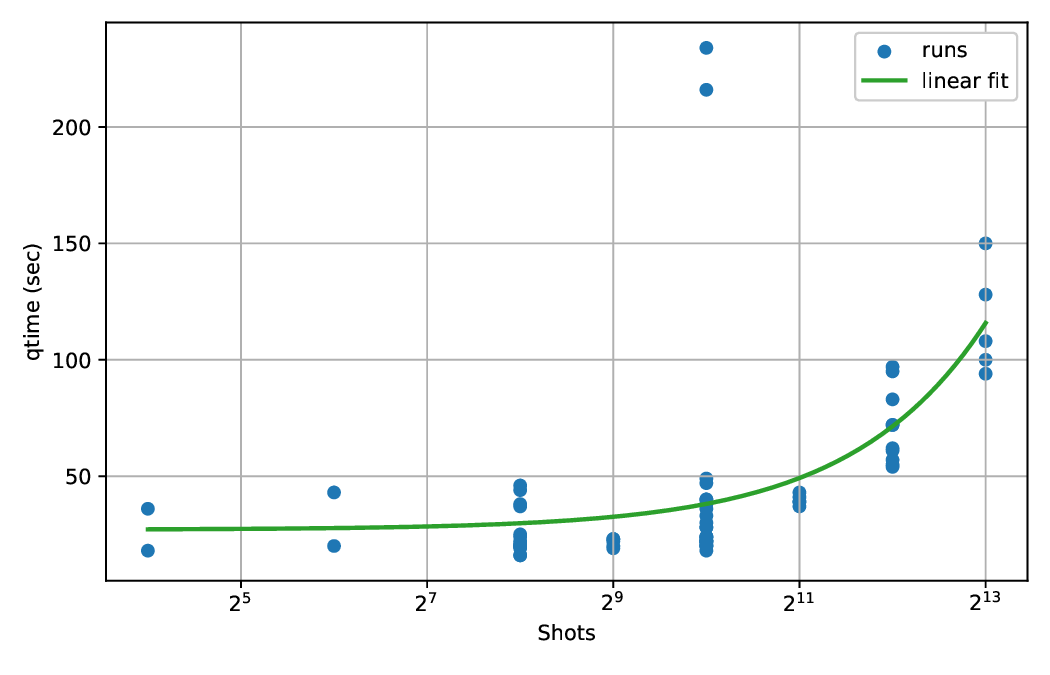}
    \includegraphics[width=0.49\linewidth]{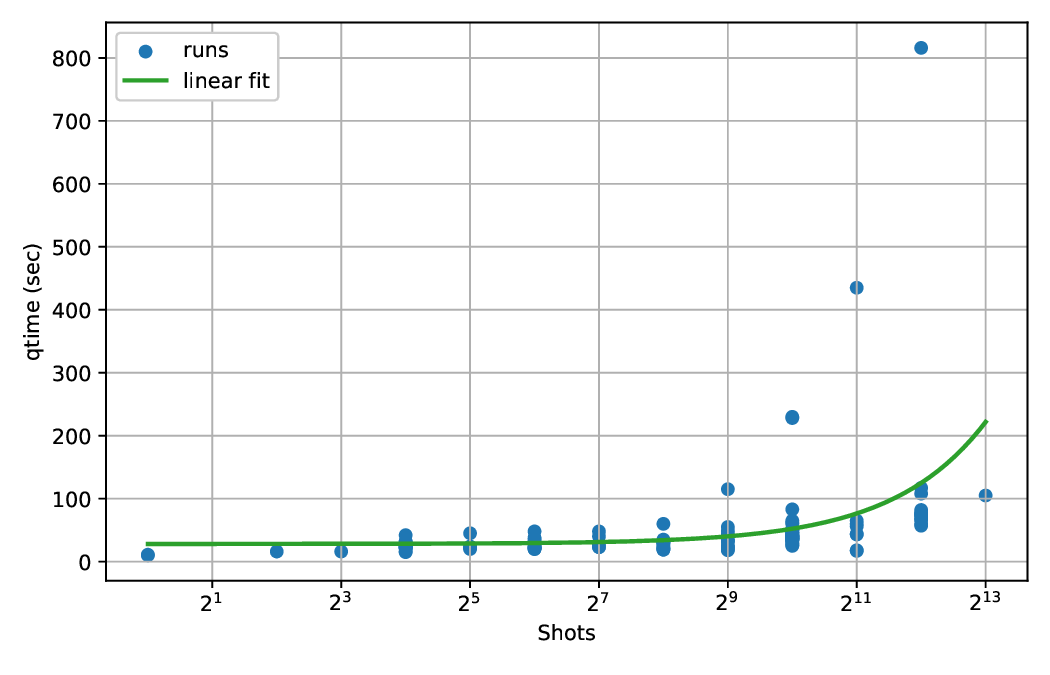}
    \caption{Quantum time as function of number of shots aggregated for all backends for PT mapper, resilience level 0, session mode (left) and single job mode (right).}
    \label{fig:qtime_vs_shots_fits}
\end{figure}

Figure~\ref{fig:13_shotsswipe_compact_all_backends_single_job} shows the equivalent to figure~\ref{fig:13_shotsswipe_compact_all_backends_session} for single job mode. And figure~\ref{fig:qtime_vs_shots_fits} shows the quantum time displayed in figures~\ref{fig:13_shotsswipe_compact_all_backends_session} and \ref{fig:02_backends_single_job} with linear fits.

\paragraph{Resilience levels and error mitigation}
\begin{table}[htbp]
\centering
\footnotesize
\setlength{\tabcolsep}{2.5pt}
\caption{Resilience impact on $E_{err}$ (PT, COBYLA, 1024 shots, session). Compact symbols are used; definitions are provided below the table.}
\label{tab:03_resilience_energy_session}
\begin{tabular}{lrrrrr}
\toprule
Backend & res & $\overline{E_{err}}$ & $E_{err}^{min}$ & $E_{err}^{max}$ & $n$ \\
\midrule
ibm\_aachen & 0 & 0.2406 & 0.1789 & 0.3548 & 4 \\
ibm\_aachen & 1 & 0.06715 & 0.0453 & 0.08901 & 2 \\
ibm\_aachen & 2 & 0.3689 & 0.2557 & 0.4822 & 2 \\
\midrule
ibm\_fez & 0 & 0.1676 & 0.09484 & 0.2665 & 3 \\
ibm\_fez & 1 & 0.1617 & 0.01105 & 0.3124 & 2 \\
ibm\_fez & 2 & 0.1213 & 0.1213 & 0.1213 & 1 \\
\midrule
ibm\_kingston & 0 & 0.1302 & 0.06519 & 0.1733 & 3 \\
ibm\_kingston & 1 & 0.00743 & 0.00743 & 0.00743 & 1 \\
ibm\_kingston & 2 & 0.0882 & 0.006862 & 0.1695 & 2 \\
\midrule
ibm\_pittsburgh & 0 & 0.2496 & 0.00549 & 0.6671 & 5 \\
ibm\_pittsburgh & 1 & 0.04383 & 0.04383 & 0.04383 & 1 \\
ibm\_pittsburgh & 2 & 0.08597 & 0.08597 & 0.08597 & 1 \\
\midrule
ibm\_brussels & 0 & 0.3508 & 0.2538 & 0.586 & 5 \\
ibm\_brussels & 1 & 0.0259 & 0.0259 & 0.0259 & 1 \\
ibm\_brussels & 2 & 0.08712 & 0.08712 & 0.08712 & 1 \\
\bottomrule
\end{tabular}
\end{table}

Table~\ref{tab:03_resilience_energy_session} reports the aggregated statistics for this configuration. Header notation: overline=mean, min/max=extrema, and n=sample count. $E_{err}=|E-E_{exact}|$ (reported as $E_{err}\times 10$ in a.u.); $btime$ is billed time (minutes unless stated otherwise); $qtime$ is quantum execution time (seconds); $res$ is resilience level.

\begin{table}[htbp]
\centering
\footnotesize
\setlength{\tabcolsep}{2.5pt}

\label{tab:03_resilience_time_session}
\begin{tabular}{lrrrrr}
\toprule
Backend & res & $\overline{btime}$ (min) & $min(btime)$ (min) & $max(btime)$ (min) & $n$ \\
\midrule
ibm\_aachen & 0 & 2.042 & 1.806 & 2.273 & 4 \\
ibm\_aachen & 1 & 5.558 & 5.277 & 5.839 & 2 \\
ibm\_aachen & 2 & 5.811 & 5.467 & 6.156 & 2 \\
ibm\_fez & 0 & 2.678 & 2.483 & 2.983 & 3 \\
ibm\_fez & 1 & 5.775 & 5.633 & 5.917 & 2 \\
ibm\_fez & 2 & 6.233 & 6.233 & 6.233 & 1 \\
ibm\_kingston & 0 & 2.868 & 2.35 & 3.422 & 3 \\
ibm\_kingston & 1 & 6.133 & 6.133 & 6.133 & 1 \\
ibm\_kingston & 2 & 6.842 & 6.533 & 7.15 & 2 \\
ibm\_pittsburgh & 0 & 2.49 & 2.15 & 3 & 5 \\
ibm\_pittsburgh & 1 & 5.733 & 5.733 & 5.733 & 1 \\
ibm\_pittsburgh & 2 & 6.567 & 6.567 & 6.567 & 1 \\
ibm\_brussels & 0 & 1.917 & 1.717 & 2.15 & 5 \\
ibm\_brussels & 1 & 5.167 & 5.167 & 5.167 & 1 \\
ibm\_brussels & 2 & 6.133 & 6.133 & 6.133 & 1 \\
\bottomrule
\end{tabular}
\caption{Impact of resileince level on billed time $btime$ in minutes for PT, COBYLA, 1024 shots, session mode. Header notation: overline=mean, min/max=extrema, and n=sample count, $res$ is resilience level.}
\end{table}

%Table~\ref{tab:03_resilience_time_session} reports the aggregated statistics for this configuration. 

\end{document}